\documentclass[onecolumn,showpacs,11pt,nofootinbib]{revtex4-2}
\usepackage{amsmath, graphicx}
\usepackage{dcolumn}
\usepackage{bm}
\usepackage{color}
\usepackage{epstopdf}
\usepackage{eurosym}
\usepackage{multirow}
\usepackage{amsfonts}
\usepackage{amsmath}
\usepackage{hyperref}
\usepackage{amssymb}
\usepackage[english]{babel}
\usepackage{graphicx}
\usepackage{epsfig}
\usepackage{subfigure}
\usepackage{unicode}
\begin{document}
\newcommand{\hs}{\hspace*{0.2cm}}
\newcommand{\hsp}{\hspace*{0.5cm}}
\newcommand{\vs}{\vspace*{0.5cm}}
\newcommand{\be}{\begin{equation}}
\newcommand{\ee}{\end{equation}}
\newcommand{\bea}{\begin{eqnarray}}
\newcommand{\eea}{\end{eqnarray}}
\newcommand{\ben}{\begin{enumerate}}
\newcommand{\een}{\end{enumerate}}
\newcommand{\bde}{\begin{widetext}}
\newcommand{\ede}{\end{widetext}}
\newcommand{\nn}{\nonumber}
\newcommand{\crn}{\nonumber \\}
\newcommand{\Tr}{\mathrm{Tr}}
\newcommand{\non}{\nonumber}
\newcommand{\noi}{\noindent}
\newcommand{\al}{\alpha}
\newcommand{\la}{\lambda}
\newcommand{\bet}{\beta}
\newcommand{\ga}{\gamma}
\newcommand{\va}{\varphi}
\newcommand{\om}{\omega}
\newcommand{\pa}{\partial}
\newcommand{\+}{\dagger}
\newcommand{\fr}{\frac}
\newcommand{\sq}{\sqrt}
\newcommand{\bc}{\begin{center}}
\newcommand{\ec}{\end{center}}
\newcommand{\Ga}{\Gamma}
\newcommand{\de}{\delta}
\newcommand{\De}{\Delta}
\newcommand{\ep}{\epsilon}
\newcommand{\varep}{\varepsilon}
\newcommand{\ka}{\kappa}
\newcommand{\La}{\Lambda}
\newcommand{\si}{\sigma}
\newcommand{\Si}{\Sigma}
\newcommand{\ta}{\tau}
\newcommand{\up}{\upsilon}
\newcommand{\Up}{\Upsilon}
\newcommand{\ze}{\zeta}
\newcommand{\ps}{\psi}
\newcommand{\Ps}{\Psi}
\newcommand{\ph}{\phi}
\newcommand{\vph}{\varphi}
\newcommand{\Ph}{\Phi}
\newcommand{\Om}{\Omega}

\newcommand{\Red}[1]{{\color{red}#1}}
\newcommand{\Revised}[1]{{\color{blue}#1}}
\newcommand{\Blue}[1]{{\color{blue}#1}}
\newcommand{\Long}[1]{{\color{cyan}#1}}
\newcommand{\Vien}[1]{{\color{red}#1}}
\title{$\mathbf{B-L}$ model with $\mathbf{D_4\times Z_4\times Z_2}$ symmetry \\ for fermion mass hierarchies and mixings}

\author{V. V. Vien}
\email{vvvien@ttn.edu.vn}
\affiliation{Department of Physics, Tay Nguyen University, 
Daklak, 
Vietnam.}

\date{\today}

\begin{abstract}
We construct a gauge $B-L$ model with $D_4\times Z_4\times Z_2$ symmetry that can explain the quark and lepton mass hierarchies and their mixings with the realistic CP phases via the type-I seesaw mechanism. Six quark mases, three quark mixing angles and CP phase in the quark sector can get the central values 
and Yukawa couplings in the quark sector are diluted a range of three orders of magnitude difference by the perturbation theory at the first order.
For neutrino sector, the smallness of neutrino mass is achieved by the Type-I seesaw mechanism. Both inverted and normal neutrino mass hierarchies are in consistent with the experimental data. The prediction for the sum of neutrino masses for normal
and inverted hierarchies, the effective neutrino masses and the Dirac CP phase are well consistent with all the recent limits.
\end{abstract}
\maketitle

\section{\label{intro} Introduction}
The mass hierarchy problem is one of the most exciting issues in particle physics that require the extension of the Standard Model (SM). Some of the experimental 
data related to flavour problem including 
the origin of the quark mass hierarchy \cite{PDG2022}
$m_u \ll m_c \ll m_t$ and 
$m_d \ll m_s \ll m_b$, the hierarchy of charged lepton mass $m_e \ll m_\mu \ll m_\tau$ and the origin of the 
tiny of three quark mixing angles as well as the neutrino mass spectrum and mixings.

Because of mentioned issues, 
various SM extensions have been implemented such as symmetry extensions with scalars and/or fermion fields. The $B-L$ model \cite{U1X0,U1X9,U1X13,U1X18,U1X6,U1X7,YFZ20} is appreciated because the simplest way is to add three right-handed neutrinos for generating neutrino masses. Although this model solves many interesting problems such as dark matter \cite{U1X9}, the muon anomalous magnetic moment \cite{U1X13,YFZ20}, leptogenesis \cite{U1X6, U1X18} and gravitational wave radiation \cite{U1X7}, it cannot provide a satisfactory explanation for fermion masses and mixing observables. Non-Abelian discrete symmetries have seem to be the most powerful tool for reproducing the observed mass and mixing patterns of leptona and quarks (see, for example, Ref. \cite{VienQ6NPB20}). $D_4$ symmetry received much attention because it can provide a predictive depiction of the mentioned patterns \cite{D41, D42, D45,VLKD4MPLA21,D46,D49v,VienD4JPG20, D47, vlD48, D4Antonioetal2021, D410vl,ACU2020,Das19,Srivastava21,D410vl, ACU2020}, however, those previous works are essentially different from our current study for the following basic points:
\begin{itemize}
\item [(1)] Ref. \cite{VienD4JPG20} based on symmetries\footnote{$G_{BL}=SU(3)_C\times SU(2)_L\times U(1)_Y\times U(1)_{B-L}$ is the gauge symmetry of $B-L$ model.} $G_{BL}\times D_4\times Z_4$ in which, for the quark sector, up to four $SU(2)_L$ doublets and three singlets are introduced, and the obtained quark mixing matrix, whose "13", "23", "31" and "32" entries are zero, is not natural because in fact all the elements of the quark mixing matrix are non-zero \cite{PDG2022}.
\item [(2)] Ref. \cite{D47} based on symmetries\footnote{$G_{SM}=SU(3)_C\times SU(2)_L\times U(1)_Y$ is the SM gauge symmetry.} $G_{SM}\times D_4\times Z_2$ in which the realistic quark mixing pattern 
    has not been considered and the quark mass hierarchy is not satisfied.
\item [(3)] Ref. \cite{vlD48} based on symmetries
    $G_{331}\times U(1)_{\mathcal{L}}\times D_4$ in which five $SU(3)_L$ triplets are used, 
    and 
    the $1-2$ mixing of the
ordinary quarks is obtained if the $D_4$ symmetry is violated with $1^'$ symmetry instead of $\underline{1}$ as usual.
\item [(4)] Ref. \cite{D410vl} based on symmetries $G_{331}\times U(1)_{\mathcal{L}}\times D_4$ in which the realistic quark mixing matrix is achievedsatisfied, however, the quark mass hierarchy is not satisfied.
\item [(5)] In Ref. \cite{Das19}, the obtained quark mixing matrix, whose "13", "23", "31" and "32" entries are zero, is not natural because in fact all the elements of the quark mixing matrix are non-zero \cite{PDG2022}, and the quark mass hierarchy is not satisfied.
\item [(6)] In Ref. \cite{ACU2020}, the obtained quark mixing matrix, whose "13", "23", "31" and "32" entries are zero, is not natural because in fact all the elements of the quark mixing matrix are non-zero \cite{PDG2022}, and the quark mass hierarchy is not satisfied.
\item [(7)] Ref. \cite{D4Antonioetal2021} based on symmetries $G_{331}\times D_{4}\times Z_{4}\times Z_{3}^{(1) } \times Z_{3}^{(2)} \times Z_{16}$ in which two $SU(3)_L$ triplets and six $SU(3)_L$ singlets are used.
\item [(8)] In Ref. \cite{Srivastava21}, the quark mass hierarchy is a bit unnatural since the Yukawa couplings spread over the region from  $\mathcal{O} (10^{-3})$ to $\mathcal{O} (1)$ (three orders of magnitude difference).
\end{itemize}
Hence, it would be desirable to propose another $D_4$ flavor model which can overcome the mentioned limitations of previous studies, especially the quark mass hierarchy, the tiny of quark mixing angles, the neutrino mass spectrum and mixing pattern.

In this study, we propose another $D_4$ model, which differs from those of Refs. \cite{VienD4JPG20, VLKD4MPLA21}, by additionally introducing one 
doublet ($H^'$) put in $\underline{1}^'$ under $D_4$ \cite{VLKD4MPLA21} and using one singlet instead of one doublet in the quark sector \cite{VienD4JPG20}. 
The properties under $D_4$ of the right handed charged lepton ($l_{1R}$) and of the right handed neutrino ($\nu_{1R}$), and the properties under $Z_4$ of right-hand leptons $l_{1R},\, l_{\alpha R}, \, \nu_{1R}, \, \nu_{\alpha R}$ and singlet scalars $\chi,\, \varphi, \, \phi$ in our present work are completely different from those of Ref. \cite{VienD4JPG20, VLKD4MPLA21}. As a consequence, the charged-leptons, neutrinos and quarks mass hierarchies can be naturally achieved. 

The rest of this work is layout as follows. We present the model description in section \ref{model}. Sections \ref{quark} and \ref{lepton} are devoted to the quark and lepton masses and mixings, respectively. Section \ref{NR} is for the numerical analysis. We make some conclusions in Sec. \ref{conclusion}.
\section{The model \label{model}}
The total symmetry of the model is $\Gamma=SU(2)_L\times U(1)_Y \times U(1)_{B-L}\times D_4\times Z_4\times Z_2$ where lepton, quark and scalar fields, under $D_4$ and $Z_4$, are essentially different from those of Refs. \cite{VienD4JPG20, VLKD4MPLA21}. Namely, in this study, the first families of the left handed quark, right handed up-and down quarks are assigned in $\mathbf{1}_{+-}$; the two other families of quarks are assigned in $\mathbf{2}$. To explain the hierarchies of quark masses, one $SU(2)_L$ doublet $H^'$ with $B-L=0$ put in $\mathbf{1}_{-+}$ under $D_4$ together with three flavons $\rho, \varphi$ and $\phi$ with $B-L=0$ respectively put in $\mathbf{2}$ and $\mathbf{1}_{+-}$ under $D_4$ are additional introduced, i.e., the considered model contains two $SU(2)_L$ doublets\footnote{see, for instance \cite{Branco12, Wang23}, for a review of the two-Higgs-doublet model (2HDM).}.
The particle and scalar contents of the model is shown 
in Table \ref{partcont}.
\begin{table}[ht]
\begin{center}
\caption{\label{partcont} Particle and scalar contents of the model ($\al=2,3$).}
\vspace{0.25 cm}
\begin{tabular}{|c||c|c|c|c|c|c||c|c|c|c|c|c||c|c|c|c|c|c|c|c|c|c|c|c|c|c|c|}
\hline
Fields &$Q_{1 L}$&$Q_{\al L}$&$u_{1 R}$&$u_{\al R}$&$d_{1R}$&$d_{\al R}$ &$\psi_{1 L}$ &$\psi_{\alpha L}$ & $l_{1 R}$ &$l_{\alpha R}$ & $\nu_{1 R}$&$\nu_{\al R}$&$H$\, &$H^'$&$\rho$ &$\phi$&$\varphi$&$\chi$\\ \hline
U(1)$_{B-L}$ &$\frac{1}{3}$&$\frac{1}{3}$&$\frac{1}{3}$&$\frac{1}{3}$&$\frac{1}{3}$&$\frac{1}{3}$& $-1$    &  $-1$    & $-1$     & $-1 $   &$-1$  &$-1$ &$0$&$0$&$0$&$0$&$0$&$2$\\
$D_4$ &$\mathbf{1}_{+-}$&$\mathbf{2}$ &$\mathbf{1}_{+-}$&  $\mathbf{2}$&$\mathbf{1}_{+-}$&$\mathbf{2}$& $\mathbf{1}_{-+}$& $\mathbf{2}$  & $\mathbf{1}_{-+}$ & $\mathbf{2}$&$\mathbf{1}_{-+}$& $\mathbf{2}$&$\mathbf{1}_{+-}$&$\mathbf{1}_{-+}$&$\mathbf{2}$&$\mathbf{1}_{+-}$&$\mathbf{1}_{--}$&$\mathbf{1}_{+-}$\\
$Z_4$    &$1$&$i$&$-1$&$-i$&$-1$&$-i$& $1$   & $i$  & $-1$ & $-i$ & $i$& $i$&$-1$&$-1$&$-i$&$1$&$-1$&$-1$\\
$Z_2$    &$-$&$+$&$-$&$+$&$-$&$+$& $+$   & $+$  & $+$ & $+$ & $-$& $-$&$+$&$+$&$-$&$+$&$-$&$+$\\
 \hline
\end{tabular}
\end{center}
\end{table} 

With the given particle content, $\bar{Q}_{1 L} u_{1 R}$ transforms as $(\textbf{2}, \frac{1}{2}, 0, \mathbf{1}_{++}, -1)$ can couple to $(\widetilde{H}\phi)_{\mathbf{1}_{++}}$; $\overline{Q}_{\al L} u_{\al R}\sim (\textbf{2}, \frac{1}{2}, 0, \mathbf{1}_{+-}+\mathbf{1}_{-+}+\mathbf{1}_{++}+\mathbf{1}_{--}, 1)$ can, respectively, couple to $\widetilde{H}, \widetilde{H^'}, (\widetilde{H} \phi)_{\mathbf{1}_{++}}$ and $(\widetilde{H^'}\phi)_{\mathbf{1}_{--}} $; $\overline{Q}_{1 L} u_{\al R} \sim (\textbf{2}, \frac{1}{2}, 0, \mathbf{2}, i)$ can couple to $(\widetilde{H}\rho)_{\mathbf{2}}$ and $(\widetilde{H^'}\rho)_{\mathbf{2}}$; and $ \overline{Q}_{\al L} u_{1 R} \sim (\textbf{2}, \frac{1}{2}, 0, \mathbf{2}, -i)$ can couple to $(\widetilde{H}\rho^*)_{\mathbf{2}}$ and $(\widetilde{H^'}\rho^*)_{\mathbf{2}}$ to form invariant terms that generate up-quark mass matrix. The situation is similar to the down quark sector.
The Yukawa terms in the quark and lepton sectors are:
\bea
-\mathcal{L}_Y^{q } &=&\frac{x_{1}^{u}}{\Lambda} (\overline{Q}_{1 L}u_{1 R})_{\mathbf{1}_{++}} (\widetilde{H}\phi)_{\mathbf{1}_{++}}
+ x_{2}^{u} (\overline{Q}_{\al L}u_{\al R})_{\mathbf{1}_{+-}} \widetilde{H}
+ x_{3}^{u}(\overline{Q}_{\al L}u_{\al R})_{\mathbf{1}_{-+}} \widetilde{H^'} \crn
&+&\frac{y_{1}^{u}}{\Lambda}(\overline{Q}_{\al L}u_{\al R})_{\mathbf{1}_{++}} (\widetilde{H}\phi)_{\mathbf{1}_{++}}
+ \frac{y_{2}^{u}}{\Lambda}(\overline{Q}_{\al L}u_{\al R})_{\mathbf{1}_{--}} (\widetilde{H^'}\phi)_{\mathbf{1}_{--}}
+ \frac{z_{1}^{u }}{\Lambda} (\overline{Q}_{1L} u_{\al R})_{\underline{2}}(\widetilde{H}\rho)_{\underline{2}}\crn
&+&\frac{z_{2}^{u }}{\Lambda} (\overline{Q}_{\al L} u_{1 R})_{\underline{2}}(\widetilde{H}\rho^*)_{\underline{2}}
+\frac{z_{3}^{u }}{\Lambda} (\overline{Q}_{1L} u_{\al R})_{\underline{2}}(\widetilde{H}^'\rho)_{\underline{2}}
+\frac{z_{4}^{u }}{\Lambda} (\overline{Q}_{\al L} u_{1 R})_{\underline{2}}(\widetilde{H}^'\rho^*)_{\underline{2}}  \crn
&+&\frac{x_{1}^{d}}{\Lambda} (\overline{Q}_{1 L}d_{1 R})_{\mathbf{1}_{++}} (H\phi)_{\mathbf{1}_{++}}
+ x_{2}^{d}(\overline{Q}_{\al L}d_{\al R})_{\mathbf{1}_{+-}} H
+ x_{3}^{d}(\overline{Q}_{\al L}d_{\al R})_{\mathbf{1}_{-+}} H^' \crn
&+&\frac{y_{1}^{d}}{\Lambda}(\overline{Q}_{\al L}d_{\al R})_{\mathbf{1}_{++}} (H \phi)_{\mathbf{1}_{++}}
+ \frac{y_{2}^{d}}{\Lambda}(\overline{Q}_{\al L}d_{\al R})_{\mathbf{1}_{--}} (H^' \phi)_{\mathbf{1}_{--}}
+ \frac{z_{1}^{d}}{\Lambda}(\overline{Q}_{1L} d_{\al R})_{\underline{2}}(H \rho)_{\underline{2}}\crn
&+& \frac{z_{2}^{d}}{\Lambda} (\overline{Q}_{\al L} d_{1 R})_{\underline{2}}(H\rho^*)_{\underline{2}}
+\frac{z_{3}^{d}}{\Lambda}(\overline{Q}_{1L} d_{\al R})_{\underline{2}}(H^' \rho)_{\underline{2}}
+ \frac{z_{4}^{d}}{\Lambda} (\overline{Q}_{\al L} d_{1 R})_{\underline{2}}(H^'\rho^*)_{\underline{2}}  + \mathrm{H.c,}
\label{Lyquark}\\
 -\mathcal{L}^{Y}_{lep}&=&\frac{h_1}{\Lambda}(\overline{\psi}_{1L}  l_{1R})_{\mathbf{1}_{++}} (H \phi)_{\mathbf{1}_{++}}
 +h_2(\overline{\psi}_{\alpha L}  l_{\alpha R})_{\mathbf{1}_{+-}} H
 +h_3(\overline{\psi}_{\alpha L}  l_{\alpha R})_{\mathbf{1}_{-+}} H^'  \crn
 &+& \frac{h_4}{\Lambda}(\overline{\psi}_{\alpha L}  l_{\alpha R})_{\mathbf{1}_{++}} (H \phi)_{\mathbf{1}_{++}}
 +\frac{h_5}{\Lambda}\big(\overline{\psi}_{\alpha L}  l_{\alpha R}\big)_{\mathbf{1}_{--}} (H^' \phi)_{\mathbf{1}_{--}} \crn
 &+&\frac{x_1}{\Lambda} \left(\bar{\psi}_{1 L}\nu_{\al  R}\right)_{\mathbf{2}} \big(\widetilde{H} \rho^*\big)_{\mathbf{2}}  +\frac{x_2}{\Lambda} \left(\bar{\psi}_{1 L}\nu_{\al  R}\right)_{\mathbf{2}} \big(\widetilde{H^'} \rho^*\big)_{\mathbf{2}} \crn
 &+&\frac{x_3}{\Lambda} \left(\bar{\psi}_{\al L}\nu_{\al R}\right)_{\mathbf{1}_{-+}} \big(\widetilde{H} \varphi\big)_{\mathbf{1}_{-+}}
 +\frac{x_4}{\Lambda} \left(\bar{\psi}_{\al L}\nu_{\al R}\right)_{\mathbf{1}_{+-}} \big(\widetilde{H^'} \varphi\big)_{\mathbf{1}_{+-}}\crn
 &+& \frac{y_1}{2\Lambda} \left(\bar{\nu}^c_{1 R}\nu_{1 R}\right)_{\mathbf{1}_{++}}\big(\phi\chi)_{\mathbf{1}_{++}}
 + y_2 (\bar{\nu}^c_{\al R} \nu_{\al R})_{\mathbf{1}_{+-}} \chi
 +\frac{y_3}{2\Lambda} (\bar{\nu}^c_{\al R}\nu_{\al R})_{\mathbf{1}_{++}} (\phi\chi)_{\mathbf{1}_{++}}
+\mathrm{H.c}, \label{Llep}\eea
where $x^{u,d}_{1,2,3}, y^{u,d}_{1,2}$ and $z^{u,d}_{1,2,3,4}$ are the Yukawa-like couplings in the quark sector, $h_{1,2,3,4,5}; x_{1,2,3,4}$ and $y_{1,2,3}$ are the Yukawa-like couplings in the lepton sector and $\Lambda$ is the cut-off scale of the theory.

It is worthy to note that additional discrete symmetries $D_4$, $Z_4$ and $Z_2$ play crucial roles in forbidding undesired terms to get the expected quark and lepton mass matrices which are listed in Table \ref{preventedterm}. For instance, in the absence of $Z_2$, there will be additional invariant terms $(\overline{\psi}_{1L} l_{\al R})_{\mathbf{2}} (H\rho)_{\mathbf{2}}, (\overline{\psi}_{1L} l_{\al R})_{\mathbf{2}} (H^'\rho)_{\mathbf{2}}, (\overline{\psi}_{\al L} l_{1 R})_{\mathbf{2}} (H\rho)_{\mathbf{2}}$ and $(\overline{\psi}_{\al L} l_{1 R})_{\mathbf{2}} (H^'\rho)_{\mathbf{2}}$ which contribute to the entries "12", "13", "21" and "31" of the charged lepton matrix. As a result, we cannot obtain the mass of charged leptons as expected since the charged lepton matrix cannot be diagonalized.

The vacuum expectation value (VEV) of the scalar fields get the form:
\bea
&&\langle H \rangle = (0 \hs \hs v)^T,\hs \langle H^' \rangle = (0 \hs \hs v^')^T, \hs \langle \varphi \rangle = v_\varphi, \hs \langle \phi \rangle = v_\phi, \crn
&& \langle \rho \rangle =\left(\langle \rho_1 \rangle, \hs  \langle \rho_1 \rangle \right)\equiv \left(v_\rho, \hs v_\rho\right), \hs \langle \chi \rangle =v_\chi. \label{scalarvev}
\eea

In fact, the electroweak symmetry breaking scale is of order about one hundred GeV, $v^2+v^{'2} =(174 \, \mathrm{GeV})^2$. Furthermore, in the 2HDM, the limits of the parameter $t_\beta =\frac{v^'}{v}$ are given by \cite{Heinemeyerprd23} $t_\beta =\frac{v^'}{v}\in [1.0. 10.0]$ or \cite{Azevedo23} $t_\beta =\frac{v^'}{v}\in [1.0. 3.0]$. For the purpose of determining the scale of Yukawa couplings, we consider the case of $t_\beta=1.424$, i.e.,
\bea
&&v=100\, \mathrm{GeV}, \hs v^'=142.40\, \mathrm{GeV}. \label{vev2H}
\eea
In addition, in order to satisfy the quark mass hierarchy, the VEV of singlets and the cut-off scale are assumed to be as follows
\bea
&&v_\rho =5\times 10^{11} \, \mathrm{GeV}, \hs v_\phi=10^{11} \, \mathrm{GeV},\hs \Lambda \simeq 10^{13}\, \mathrm{GeV}. \label{vevscales}
\eea
The models with more than one $SU(2)_L$ 
scalar doublet as in this work, the Flavor Changing Neutral Current (FCNC) processes such as $b\rightarrow s \gamma$ exist in the Higgs sector. However, they are suppressed by non-Abelian discrete symmetries \cite{Mondragon07,Kubo13}. To make such process below the current experimental limits, some restrictions on the model parameters such as the Yukawa couplings and large masses for non SM scalars need to be imposed. The considered model contains many free parameters which allows us freedom to assume that the remaining scalars are sufficiently heavy to fullfil the current experimental limits.
Furthermore, the first two lines of Eq. (\ref{Llep}) imply that the off-diagonal Yukawa couplings in the charged-lepton sector are proportional to $\frac{v_{\phi}}{\La}\sim 10^{-2}$. Therefore, the lepton flavor violation (LFV) processes, such as $l_j\rightarrow l_i \gamma$, are suppressed by the tiny factor $\fr{v_{\phi}}{\La}\fr{1}{m_H^2}$ associated
with the mentioned small Yukawa couplings and the large mass scale of the heavy scalars $m_H$ \cite{Dorsner15, Davidson10, Davidson16,Vien2021}. A detailed study of FCNC and LFV processes are beyond the scope of this work.
\section{\label{quark} Quark mass and mixing}
Using the Clebsch-Gordan coefficients of $D_4$ symmetry \cite{Ishi}, from Eq. (\ref{Lyquark}), when the scalar fields get the VEVs as, Eq. (\ref{scalarvev}), 
the up-and down-quark mass matrices take the following forms:
\bea
M_{q} =M^{(0)}_{q} + \delta M_{q} \hs\,\, (q=u, d), \label{Mq}\eea
where
\bea
&&M^{(0)}_{q}=\left(
\begin{array}{ccc}
a_{1q} &0 &0 \\
0 &a_{2q}+a_{3q}& 0 \\
0 &0& a_{2q}- a_{3q}
\end{array}%
\right), \hs  \delta M_{q}=\left(
\begin{array}{ccc}
0 & c_{1q}+c_{3q}& c_{1q}-c_{3q}  \\
c_{2q}+c_{4q} &0& b_{1q}+b_{2q} \\
c_{2q}-c_{4q} &b_{1q}- b_{2q} & 0
\end{array}%
\right), \label{M0delMq}
\eea
with
\bea
&&a_{1q}=x_{1}^{q} v \frac{v_\varphi}{\Lambda}, \hs a_{2q}=x_{2}^{q} v, \hs a_{3q}=x_{3}^{q} v^', \hs
b_{1q}=y_{1}^{q} v\frac{v_\phi}{\Lambda}, \hs b_{2q}=y_{2}^{q} v^'\frac{v_\phi}{\Lambda},\crn
&&c_{1q}=z_{1}^{q} v\frac{v_\rho}{\Lambda},\hs c_{2q}=z_{2}^{q} v\frac{v_\rho}{\Lambda}, \hs c_{3q}=z_{3}^{q} v^'\frac{v_\rho}{\Lambda}, \hs c_{4q}=z_{4}^{q} v^'\frac{v_\rho}{\Lambda} \hs\, (q=u, d). \label{abcud}
\eea
Expressions (\ref{Mq})$-$(\ref{abcud}) show that, besides two doublets $H$ and $H^'$, one singlet $\varphi$ contributes to $M^{(0)}_{q}$ while $\delta M_{q}$ is due to the contribution of two singlets $\rho$ and $\phi$. Without the contributions of $\rho$ and $\phi$, $\delta M_{q}$ will be vanished and the quark mass matrices $M_{q}$ in Eq. (\ref{Mq}) reduce to the diagonal matrices $M^{(0)}_{q}$, i.e., the corresponding quark mixing matrix $V_{CKM}=\mathbb{I}_{3\times 3}$ which was ruled out by the recent data. The realistic quark mixing angles are very small \cite{PDG2022} which implies that the quark mixing matrix is very close to the identity matrix; thus, the second term $\delta M_{q}$ in Eq.(\ref{M0delMq}) can be considered as the perturbed parameter for generating the quark mixing pattern. As a consequence, the realistic quark mixing pattern can be achieved at the first order of perturbation theory. Indeed, at the first order of perturbed theory, the matrices
$\delta M_{q}$ contribute to the eigenvectors but they have no contribution to the eigenvalues of the quark mass matrices $M_q$. The 
quark masses are determined as
\bea m_{u} &=&a_{1u},\hs m_{c}=a_{2u} + a_{3u},\hs m_t=a_{2u} - a_{3u}, \crn
 m_{d} &=&a_{1d},\hs m_{s}=a_{2d} + a_{3d},\hs m_b=a_{2d} - a_{3d}, \label{quarkmasses}\eea
and the corresponding perturbed
quark mixing matrices are:
 \bea
 &&\hspace{-1.0 cm}U^{u}_{L}=U^{u}_{R}=\left(
\begin{array}{ccc}
 1 & \frac{c_{1u}+c_{3u}}{m_c-m_{u}} & \frac{c_{1u}-c_{3u}}{m_t-m_{u}} \\
 \frac{c_{4u}+c_{2u}}{m_{u}-m_c} & 1 & \frac{b_{2u}+b_{1u}}{m_{t}-m_c} \\
 \frac{c_{4u}-c_{2u}}{m_{t}-m_u} & \frac{b_{2u}-b_{1u}}{m_{t}-m_c} & 1 \\
\end{array}
\right),\,\,\,  
U^{d}_{L}=U^{d}_{R}=\left(
\begin{array}{ccc}
 1 & \frac{c_{1d}+c_{3d}}{m_s-m_{b}} & \frac{c_{1d}-c_{3d}}{m_b-m_{d}} \\
 \frac{c_{4d}+c_{2d}}{m_{d}-m_s} & 1 & \frac{b_{2d}+b_{1d}}{m_{b}-m_s} \\
 \frac{c_{4d}-c_{2d}}{m_{b}-m_d} & \frac{b_{2d}-b_{1d}}{m_{b}-m_s} & 1 \\
\end{array}
\right), \label{UudLR}
 \eea
with $b_{1,2 q}$ and $c_{1,2,3,4 q}\, (q=u,d)$ are given in Eq. (\ref{abcud}). For simplicity, we consider tha case of $y_{1q}=y_{2q}=y_{q} \, (q=u,d), \, z_{3d}=z_{1d}=z_{d}$, i.e.,
\bea
&&b_{2d} = b_{1d} = b_d, \hs
b_{2u} = b_{1u} = b_u, \hs
c_{3d} = c_{1d}. \label{asum}
\eea
The quark mixing matrix, $V_\mathrm{CKM}= V^u_L V^{d \dagger}_L$, owns the following entries:
\bea
&&V_\mathrm{CKM}^{11}=1+\frac{2 c^*_{1d} (c_{1u}+c_{3u})}{(m_{u}-m_{c}) (m_{d}-m_{s})}, \crn
&&V_\mathrm{CKM}^{12}=\frac{2 b^*_{d} (c_{1u}-c_{3u})}{(m_{b}-m_{s}) (m_{t}-m_{u})}+\frac{c_{1u}+c_{3u}}{m_{c}-m_{u}}+\frac{c^*_{2d}+c^*_{4d}}{m_{d}-m_{s}},\crn
&& V_\mathrm{CKM}^{13}=\frac{c_{1u}-c_{3u}}{m_{t}-m_{u}}+\frac{c^*_{4d}-c^*_{2d}}{m_{b}-m_{d}},\hs\hs
V_\mathrm{CKM}^{21}=\frac{c_{2u}+c_{4u}}{m_{u}-m_{c}}+\frac{2 c^*_{1d}}{m_{s}-m_{d}}, \crn
&&V_\mathrm{CKM}^{22}=1+\frac{4 b^*_{d} b_{u}}{(m_{b}-m_{s}) (m_{t}-m_{c})}+\frac{(c_{2u}+c_{4u}) (c^*_{2d}+c^*_{4d})}{(m_{u}-m_{c}) (m_{d}-m_{s})}, \label{Vckmentries}\\
&&V_\mathrm{CKM}^{23}=\frac{2 b_{u}}{m_{t}-m_{c}}+\frac{(c_{2u}+c_{4u}) (c^*_{2d}-c^*_{4d})}{(m_{b}-m_{d}) (m_{c}-m_{u})},\hs V_\mathrm{CKM}^{31}=\frac{c_{4u}-c_{2u}}{m_{t}-m_{u}},\crn
&&V_\mathrm{CKM}^{32}=\frac{2 b^*_{d}}{m_{b}-m_{s}}+\frac{(c_{2u}-c_{4u}) (c^*_{2d}+c^*_{4d})}{(m_{d}-m_{s}) (m_{u}-m_{t})},\,\, V_\mathrm{CKM}^{33}=1 + \frac{(c_{2u}-c_{4u}) (c^*_{2d}-c^*_{4d})}{(m_{b}-m_{d}) (m_{t}-m_{u})}. \nonumber
\eea
Comparing the model results on the quark masses and quark mixing matrix in Eqs. (\ref{quarkmasses}) and (\ref{Vckmentries}) with their corresponding experimental constraints on $\mathrm{V}^{\mathrm{exp}}_{ij}$ as shown in Tab. \ref{quarkpara} (the second column), we get the explicit expressions of $a_{1u,d}, a_{2u,d}, a_{3u,d}, b_{u,d}, c_{1u, d}$, $c_{2u,d}$, $c_{3u}$ and $c_{4u,d}$ as functions of quark masses and quark mixing matrix elements as presented in Eqs. (\ref{abcudexpress}) and (\ref{FGHTKP}) of Appendix \ref{abcudexpressions}.

Expressions (\ref{abcud}), (\ref{asum}), (\ref{abcudexpress}) and (\ref{FGHTKP}) imply that the model parameters $a_{1u,d}, a_{2u,d}, a_{3u,d}, b_{u,d}, c_{1u, d}$, $c_{2u,d}$, $c_{3u}$ and $c_{4u,d}$ depend on the observed parameters in the quark sector, including quark masses $m_u, m_c,m_t,m_d,m_s,m_b$ and quark mixing matrix elements $V^{\mathrm{exp}}_{ij}\, (i,j=1,2,3)$, that have been determined accurately \cite{PDG2022}. At the best-fit points of mentioned parameters\footnote{The best-fit points in Table \ref{quarkpara} correspond to the Wolfenstain parameters\cite{PDG2022}: $\lambda=0.2250,\hs A=0.826, \hs \bar{\rho}=0.159$ and $\bar{\eta}=0.348$ which correspond to the mixing angles $\sin \theta^{q}_{12} = 0.22500, \hs \sin \theta^{q}_{13}= 0.00369, \hs \sin \theta^{q}_{23}=0.04182$ and $\delta^{q}_{CP}=1.444$.} given in Refs.\cite{PDG2022}, we obtain a prediction for the quark mixing matrix and the model's parameters in the quark sector as shown in Table \ref{quarkpara} and Eq. (\ref{modelpara}), respectively.
\begin{table}[tbh]
\caption{\label{quarkpara}The best-fit points for quark parameters taken from Ref.\cite{PDG2022} and the model prediction.}
\vspace{-0.25cm}
\begin{center}
\begin{tabular}{|c|c|c|c|}
\hline
Observable & Best-fit point \cite{PDG2022} & The model prediction& Percent error $(\%)$ \\ \hline
$m_u [\mathrm{MeV}]$  &  \quad $2.16$ &$2.16$& $0$ \\ \hline
$m_c [\mathrm{GeV}]$ & \quad $1.27$ &$1.27$& $0$ \\ \hline
$m_t [\mathrm{GeV}]$&  \quad $172.69$ &$172.69$& $0$ \\ \hline
$m_d [\mathrm{MeV}]$&  \quad $4.67$ &$4.67$& $0$ \\ \hline
$m_s [\mathrm{MeV}]$&  \quad $93.4$ &$93.4$& $0$ \\ \hline
$m_b [\mathrm{GeV}]$ &  \quad $4.18$ &$4.18$& $0$ \\ \hline
$V_{\mathrm{CKM}}^{11}$ &  \quad $0.974352$ &  \quad $0.974352$& $0$\\ \hline
$V_{\mathrm{CKM}}^{12}$ &  \quad $0.224998$ &  \quad $0.224998$& $0$\\ \hline
$V_{\mathrm{CKM}}^{13}$ &  \quad $0.0015275 - 0.003359 i$ &   $0.0015275 - 0.003359 i$& $0$ \\ \hline
$V_{\mathrm{CKM}}^{21}$ &  \quad $-0.224865 - 0.000136871 i$&  $-0.224865 - 0.000136871 i$& $0$ \\ \hline
$V_{\mathrm{CKM}}^{22}$ &  \quad $0.973492$&   $0.973492$ & $0$\\ \hline
$V_{\mathrm{CKM}}^{23}$ &  \quad $0.0418197$ &$0.0418197$& $0$\\ \hline
$V_{\mathrm{CKM}}^{31}$ &  \quad $0.00792247 - 0.00327 i$ &$0.00792247 - 0.00327 i$& $0$\\ \hline
$V_{\mathrm{CKM}}^{32}$ &  \quad $-0.0410911 - 0.000755113 i $&$-0.0410911 - 0.000755113 i $ &$0$\\ \hline
$V_{\mathrm{CKM}}^{33}$ &  \quad $0.999118$ &$0.999118$& $0$ \\ \hline
\end{tabular}%
\end{center}
\vspace{-0.5cm}
\end{table}
\bea
&&a_{1u}=2.160\times 10^{-3} \,\mathrm{GeV}, \hs
a_{2u}=86.980 \mathrm{GeV}, \hs a_{3u}=-85.710 \,\mathrm{GeV},  \crn
&&b_u=(2.308+0.5413 i) \,\mathrm{GeV}, \,\,
c_{1u}=8.414+3.028i\, \mathrm{GeV}, \crn
&&c_{2u}=(-0.614+0.211 i)\, \mathrm{GeV},\,\, c_{3u}=(-8.269-3.170 i)\, \mathrm{GeV}, \crn
&&c_{4u}=(0.754-0.353 i)\, \mathrm{GeV}, \,\, a_{1d}=4.670\times 10^{-3} \,\mathrm{GeV}, \hs a_{2d}=2.140\, \mathrm{GeV}, \crn
&&a_{3d}=-2.040 \,\mathrm{GeV}, \hs b_d=(-8.658+0.262 i)10^{-2} \,\mathrm{GeV}, \crn
&&c_{1d}=(-5.080+4.973i) 10^{-3}\, \mathrm{GeV},\hs c_{2d}=(0.193-0.077 i)\, \mathrm{GeV}, \crn
&&c_{4d}=(-0.204+0.087 i)\, \mathrm{GeV}. \label{modelpara}
\eea
The Jarlskog invariant in the quark sector, $J_{CP}^{q}= \mathrm{Im}\big[V_{us} V_{cb} V_{cs}^* V_{ub}^*\big]$, is calculated from Eq. (\ref{Vckmentries}) with the model result in Table \ref{quarkpara} (the third column) as $J_{CP}^{q}=3.08\times 10^{-5}$, which coincides with 
that of Ref. \cite{PDG2022}.

Next, comparing Eqs. (\ref{abcud}) and (\ref{modelpara}) with the aid of Eqs. (\ref{vev2H})-(\ref{vevscales}), ones obtain:
\bea
&&|x_{1u}|=2.16\times 10^{-3}, \hs |x_{2u}| =0.87, \hs |x_{3u}| =0.60,\hs |y_{1u}|=2.37, \crn
&&|y_{2u}|=1.67,\hs |z_{1u}|=1.79, \hs |z_{2u}|=0.13, \hs|z_{3u}|=1.24, \hs|z_{4u}|=0.12,\crn
&&|x_{1d}|=4.67\times 10^{-3}, \hs |x_{2d}| =2.14\times 10^{-2}, \hs |x_{3d}| =1.43\times 10^{-2},\crn
&&|y_{1d}|=8.66\times 10^{-2}, \hs |y_{2d}|=6.08\times 10^{-2}, \hs |z_{1d}|=1.42\times 10^{-3}, \crn
&&|z_{2d}|=4.16\times 10^{-2}, \hs |z_{3d}|=10^{-2}, \hs|z_{4d}|=3.11\times 10^{-2}, \label{modelpara1}
\eea
which differ by about three orders of magnitude.
\section{Lepton masses and mixings \label{lepton}}
Using the Clebsch-Gordan coefficients of $D_4$\cite{Ishi}, from Eq. (\ref{Llep}), when the scalar fields get the VEVs, Eq. (\ref{scalarvev}), we find charged leptons ($M_l$) and neutrino (Dirac and right-handed Majorana) mass matrices $(M_D, M_R)$ as follows

 \bea M_l=
\left(%
\begin{array}{ccc}
  a_1 & 0 & 0 \\
   0 &  a_2+a_3& a_4+a_5\\
  0 & a_4-a_5  & a_2-a_3\\
\end{array}%
\right),\,  M_D=\left(%
\begin{array}{ccc}
0& -a_{D}+b_{D}& a_{D}+b_{D} \\
0& c_{D}+d_{D} & 0   \\
0& 0           & -c_{D}+d_{D} \\
\end{array}%
\right),\, M_R=
\left(%
\begin{array}{ccc}
a_R& 0 & 0 \\
0  & b_R  & c_R \\
0  & c_R  & b_R \\
\end{array}%
\right),\hs  \label{MDR}\eea
where
\bea
a_1&=& \left(\frac{v_\phi }{\Lambda}\right) v h_1, \hs a_2=h_2 v, \hs a_3=h_3 v^',\hs a_4= \left(\frac{v_\phi }{\Lambda}\right) v h_4, \hs a_5= \left(\frac{v_\phi }{\Lambda}\right) v^' h_5. \label{aiexpres}\\
a_D&=&\left(\frac{v_{\rho}}{\Lambda}\right) x_1 v,\hs b_D = \left(\frac{v_{\rho}}{\Lambda}\right) x_2 v^',
\hs c_D = \left(\frac{v_{\varphi}}{\Lambda}\right) x_3 v, \hs d_D = \left(\frac{v_{\varphi}}{\Lambda}\right) x_4 v^', \crn
a_R&=& \frac{y_1}{\Lambda} v_\chi v_{\phi}, \hspace{0.55 cm}  b_R=y_2 v_\chi, \hspace{1.15 cm} c_R=\frac{y_3}{\Lambda} v_\chi v_{\phi}.\label{abcfg}\eea
$\bullet$ \emph{Charged-lepton sector:} For simplicity, we consider the case of $\arg h_3=(\arg h_2+\pi)$ and $\arg h_5=\arg h_4$, i.e, $\arg a_3=(\arg a_2+\pi)$ and $\arg a_5=\arg a_4$. Yukawa couplings $h_i \, (i=1\div 5)$ are complex in general, therefore the matrix $M_l$ is complex and its eigenvalues are complex.
Let us first define a Hermitian matrix $m^2_l=M_l M^{\+}_l$, given by
\bea
m^2_l&=&  M_l M^+_l =\left(
\begin{array}{ccc}
 A_{0} & 0 & 0 \\
 0 & B_{0} & \mathcal{D}_0. e^{-i\theta} \\
 0 & \mathcal{D}_0. e^{i\theta} & C_{0} \\
\end{array}
\right), \label{mlsq}
\eea
where\footnote{In this work, the following notations are used: $s_\psi=\sin \psi,\, c_\psi=\cos \psi, \, s_\theta=\sin \theta,\, c_\theta=\cos \theta,\, t_{\alpha}=\tan \alpha,\,t_{\theta}=\tan\theta,\, s_\delta=\sin\delta_{CP},\, s_{ij}=\sin\theta_{ij},\, c_{ij}=\cos\theta_{ij}$ and $t_{ij}=\tan\theta_{ij}\, (ij=12,13,23)$.}
\bea
&&A_0 = |a_{1}|^2, \hs
B_0 = \big(|a_{2}| - |a_{3}|\big)^2 + \big(|a_{4}| + |a_{5}|\big)^2,\hs C_0 = \big(|a_{2}| + |a_{3}|\big)^2 + \big(|a_{4}| -|a_{5}|\big)^2, \crn
&&D_0 = 2 \big(|a_{2}| |a_{4}| + |a_{3}| |a_{5}|\big) c_\alpha, \hs
G_0 = -2 \big(|a_{3}| |a_{4}| + |a_{2}| |a_{5}|\big)s_\alpha, \hs \mathcal{D}_0=\sqrt{D^2_0+G^2_0},  \label{ABCD0}\\
&&\theta=\arccos \left(\frac{D_0}{\mathcal{D}_0}\right), \hs \alpha=\arg a_2 - \arg a_4.  \label{theta}
\eea
The matrix $m^2_l$ in Eq. (\ref{mlsq}) is diagonalised by two mixing matrices $V_{l(L, R)}$ with $V^+_{lL} m^2_l V_{lR}=\mathrm{diag} (m^2_e, m^2_\mu, m^2_\tau)$, where
\bea
&&m^2_e =A_{0}, \,\, m^2_{\mu, \tau}= \frac{1}{2} \left(B_0+C_0\mp \sqrt{(B_0-C_0)^2+4 \mathcal{D}_0^2}\right),  \label{memt}\\
&&V_{lL}=V_{lR}=\left(%
\begin{array}{ccc}
  1 & 0 & 0 \\
   0 & c_\psi &\,\,\,\,\,\, -s_\psi . e^{-i \theta}\\
  0 &\hs s_\psi . e^{i \theta} & c_\psi  \\
\end{array}%
\right), \label{UClep}\eea
where
\bea
&&s_\psi
= \frac{1}{\sqrt{2}\sqrt{1-\frac{B_0-C_0}{B_0-C_0+\sqrt{(B_0-C_0)^2+4 \mathcal{D}^2_0}}}}. \label{spsi}
\eea
Equations (\ref{ABCD0})-
(\ref{memt}) and (\ref{spsi}) yield the following relations:
\bea
&&|a_{1}|=m_e, \, |a_{2}|=\frac{|a_{4}| D_0 s_\alpha+|a_{5}| c_\alpha G_0}{\big(|a_{4}|^2 - |a_{5}|^2\big) s_{2\alpha}}, \, |a_{3}| =\frac{|a_{4}| c_\alpha G_0 + |a_{5}| D_0 s_\alpha}{(|a_{5}|^2-|a_{4}|^2)s_{2\alpha}}, \crn
&&|a_{4}| = \frac{a + b}{2}, \hs |a_{5}| = \frac{a - b}{2}, \label{a0b0c0d0}\eea
where
\bea
&&a=\sqrt{\frac{\sqrt{(B_0 C_0-x_0+y_0)^2-4 B_0 C_0 y_0}+B_0 C_0-x_0+y_0}{2C_0}}, \crn
&&b=\sqrt{\frac{\sqrt{(B_0 C_0-x_0+y_0)^2-4 B_0 C_0 y_0}+B_0 C_0+x_0-y_0}{2 B_0}}, \label{abexpres}\\
&&x_0=\frac{\big(c_\alpha G_0+D_0 s_\alpha\big)^2}{s^2_{2\alpha}}, \hs y_0=\frac{\big(c_\alpha G_0-D_0 s_\alpha\big)^2}{s^2_{2\alpha}},  \label{xy0expres}\\
&&B_0=\left(m_\mu^2-m_\tau^2\right) s_\psi^2 +m_\tau^2, \hs C_0=\left(m_\mu^2-m_\tau^2\right)s_\psi^2 +m_\mu^2, \crn
&&D_0=\left(m_\tau^2-m_\mu^2\right)c_\theta s_\psi c_\psi ,\hs G_0=\left(m_\mu^2-m_\tau^2\right)s_\theta s_\psi c_\psi . \label{BCDG0expres}
\eea
Expressions (\ref{aiexpres}) and (\ref{a0b0c0d0})-(\ref{BCDG0expres}) imply that $h_{1}$ depends on $m_e, \La, v_\phi$ and $v$; $h_{2}$ depends on $v, m_\mu, m_\tau, \psi, \theta$ and $\alpha$; $h_{3}$ depends on $v^', m_\mu, m_\tau, \psi, \theta$ and $\alpha$; and $h_{4}$ and $h_5$ depend on $v, \Lambda, v_\phi, m_\mu, m_\tau$, $\psi, \theta$ and $\alpha$.
As will see in Sec. \ref{NR}, with the observed charged leptons $m_{e,\mu,\tau}$ \cite{PDG2022} and the cut-off scale, the VEV scales of scalar fields in Eqs (\ref{vev2H}) and
(\ref{vevscales}), there exist possible ranges of the model parameters such that the Yukawa couplings in the charged lepton sector, $h_{i}\, (i=1\div5)$, differ by about two orders of magnitude, i.e., the charged lepton mass hierarchy is satisfied.

$\bullet$ \emph{Neutrino sector:}
The effective neutrino mass matrix arise from type-I seesaw
mechanism $M_{\nu}=-M_D M_R^{-1} M^T_D$, obtained from Eq. (\ref{MDR}), as follows:
 \bea
M_{\nu}&=&\left(
\begin{array}{ccc}
 A & -B_1 & -B_2 \\
 -B_1 & C_1 & C_3 \\
 -B_2 & C_3 & C_2 \\
\end{array}
\right), \label{Meff}\eea
where
\bea
&&A=\frac{2 b_{D}^2}{b_{R}+c_{R}}+\frac{2 a_{D}^2}{b_{R}-c_{R}},
\hs B_{1}=\frac{(c_{D}+d_{D}) \big[a_{D} (b_{R}+c_{R})-b_{D} (b_{R}-c_{R})\big]}{b_R^2-c_R^2}, \crn
&&B_{2}=\frac{(c_{D}-d_{D}) \big[a_{D} (b_{R}+c_{R})+b_{D} (b_{R}-c_{R})\big]}{b_{R}^2-c_{R}^2},
\hs C_{1}=\frac{b_{R} (c_{D}+d_{D})^2}{b_{R}^2-c_{R}^2},  \crn
&&C_{2}=\frac{b_{R} (c_{D}-d_{D})^2}{b_{R}^2-c_{R}^2}, \hs C_{3}=\frac{c_{R} \left(c_{D}^2-d_{D}^2\right)}{b_{R}^2-c_{R}^2}.\label{abcdgh}
\eea
The mass matrix $M_{\nu}$ in Eq.(\ref{Meff}) owns three eigenvalues and the corresponding mixing matrix as follows:
\bea
&&\lambda_1=0,\hs \lambda_{2}=\frac{C_2 - 2 B_2 n_1 + A n_1^2 + n_2 (2 C_3 - 2 B_1 n_1 + C_1 n_2)}{n_{1}^2+n_{2}^2+1}, \crn
&&\lambda_{3}=\frac{C_2 - 2 B_2 t_1 + A t_1^2 + t_2 (2 C_3 - 2 B_1 t_1 + C_1 t_2)}{t_{1}^2+t_{2}^2+1},\label{la123}\\
&&\mathrm{R}=\left( \begin{array}{ccc}
\frac{k_1}{\sqrt{1 + k_1^2 (1 + k_2^2)}} &\frac{n_1}{\sqrt{n^2_1+n^2_2+1}}&\frac{t_1}{\sqrt{t^2_1+t^2_2+1}}\\
\frac{k_1 k_2}{\sqrt{1 + k_1^2 (1 + k_2^2)}}&\frac{n_2}{\sqrt{n^2_1+n^2_2+1}}&\frac{t_2}{\sqrt{t^2_1+t^2_2+1}}\\
\frac{1}{\sqrt{1 + k_1^2 (1 + k_2^2)}}  &\frac{1}{\sqrt{n^2_1+n^2_2+1}} &\frac{1}{\sqrt{t^2_1+t^2_2+1}}
\end{array}\right),\label{neumix}
\eea
where new parameters $k_{1,2}, n_{1,2}$ and $t_{1,2}$, own explicit expressions in Appendix \ref{k12n12t12expressions}, satisfy the following relations
\bea
 && k_1 (n_1 + k_2 n_2)+1=0,\,\,  k_1 (t_1 + k_2 t_2)+1=0, \,\, n_1 t_1 + n_2 t_2+1=0,  \label{knt12relat}\\
 &&C_2 - B_2 (k_1 + n_1) + C_3 (k_1 k_2 + n_2) +
 k_1 \big[A n_1 + C_1 k_2 n_2 - B_1 (k_2 n_1 + n_2)\big]=0, \label{knt12relat1}\\
 &&C_2 - B_2 (k_1 + t_1) + C_3 (k_1 k_2 + t_2) +  k_1 \big[A t_1 + C_1 k_2 t_2 - B_1 (k_2 t_1 + t_2)\big]=0, \label{knt12relat2}\\
 &&C_2 + C_3 n_2 + A n_1 t_1 - B_1 n_2 t_1 - B_2 (n_1 + t_1) + (C_3 - B_1 n_1 + C_1 n_2) t_2=0, \label{knt12relat3}\\
 &&C_2 + k_1 \big[2 C_3 k_2 -2 B_2+ k_1 (A - 2 B_1 k_2 + C_1 k_2^2)\big]=0.\label{knt12relat4}
 \eea
Depending on the sign of $\Delta m^2_{31}$, the neutrino mass spectrum can be normal or inverted hierarchy \cite{PDG2022}. In the considered model, $0=m_1\equiv \la_1< m_{2}\equiv \la_2< m_{3}\equiv \la_3$ for NH and $0=m_3\equiv \la_1< m_1\equiv \la_2< m_2\equiv \la_3$ for IH. Since the lightest neutrino mass is equal to zero, other neutrino masses and their sum are given by
\bea
&&\left\{
\begin{array}{l}
m_1=0, \hs\hs m_2=\sqrt{\Delta m^2_{21}}, \hs\hs m_3=\sqrt{\Delta m^2_{31}} \hspace{0.25cm}\mbox{for NH,}\ \  \\
m_1=\sqrt{-\Delta m^2_{31}}, \hs m_2=\sqrt{\Delta m^2_{21}-\Delta m^2_{31}}, \hs m_3=0 \hspace{0.2cm}\mbox{for IH.}
\end{array}%
\right.  \label{m1m2m3}\\
&&\sum m_\nu=\left\{
\begin{array}{l}
\sqrt{\Delta m^2_{21}}+\sqrt{\Delta m^2_{31}} \hspace{0.2cm}\mbox{for NH},    \\
\sqrt{\Delta m^2_{21}-\Delta m^2_{31}} +\sqrt{-\Delta m^2_{31}} \hspace{0.2cm}\mbox{for IH}.
\end{array}%
\right. \label{sumexpresion}
\eea
The neutrino mass matrix $M_{\nu}$ in Eq. (\ref{Meff}) is diagonalized as
 \begin{equation}
\mathrm{U}_{\nu }^T M_{\nu} \mathrm{U}_{\nu }=\left\{
\begin{array}{l}
\left(
\begin{array}{ccc}
0 & 0 & 0 \\
0 & m_{2} & 0 \\
0 & 0 & m_{3}%
\end{array}%
\right) ,\hspace{0.1cm} \mathrm{U}_{\nu }=\left( \begin{array}{ccc}
\frac{k_1}{\sqrt{1 + k_1^2 (1 + k_2^2)}} &\frac{n_1}{\sqrt{n^2_1+n^2_2+1}}&\frac{t_1}{\sqrt{t^2_1+t^2_2+1}}\\
\frac{k_1 k_2}{\sqrt{1 + k_1^2 (1 + k_2^2)}}&\frac{n_2}{\sqrt{n^2_1+n^2_2+1}}&\frac{t_2}{\sqrt{t^2_1+t^2_2+1}}\\
\frac{1}{\sqrt{1 + k_1^2 (1 + k_2^2)}}  &\frac{1}{\sqrt{n^2_1+n^2_2+1}} &\frac{1}{\sqrt{t^2_1+t^2_2+1}}
\end{array}\right) \hspace{0.2cm}\mbox{for NH,}\ \  \\
\left(
\begin{array}{ccc}
m_{1} & 0 & 0 \\
0 & m_{2} & 0 \\
0 & 0 & 0
\end{array}%
\right) ,\hspace{0.1cm} \mathrm{U}_{\nu }=\left(\begin{array}{ccc}
\frac{n_1}{\sqrt{n^2_1+n^2_2+1}}&\frac{t_1}{\sqrt{t^2_1+t^2_2+1}}&\frac{k_1}{\sqrt{1 + k_1^2 (1 + k_2^2)}}\\
\frac{n_2}{\sqrt{n^2_1+n^2_2+1}}&\frac{t_2}{\sqrt{t^2_1+t^2_2+1}}&\frac{k_1 k_2}{\sqrt{1 + k_1^2 (1 + k_2^2)}} \\
\frac{1}{\sqrt{n^2_1+n^2_2+1}} &\frac{1}{\sqrt{t^2_1+t^2_2+1}}& \frac{1}{\sqrt{1 + k_1^2 (1 + k_2^2)}}
\end{array}\right) \hspace{0.2cm}\mbox{for IH,}%
\end{array}%
\right.  \label{Unu}
\end{equation}
where $\lambda_{2}, \lambda_3$, $k_{1,2}, n_{1,2}$ and $t_{1,2}$ are given in Appendix \ref{k12n12t12expressions}.

Expressions (\ref{la123}) and (\ref{knt12relat})-(\ref{knt12relat4}) yield:
 \bea
 &&\left\{
\begin{array}{l}
k_{1}=\frac{n_{1} t_{1}+n_{2}^2+1}{t_{1} \left(n_{1}^2+n_{2}^2\right)+n_{1}},  \hs k_{2}=\frac{n_{2} (t_{1}-n_{1})}{n_{1} t_{1}+n_{2}^2+1},
 \hs t_{2}=-\frac{n_{1} t_{1}+1}{n_{2}} \hspace{1.15cm}\mbox{for NH,}\ \  \\
n_2 = \frac{1 - k_1 n_1}{k_1 k_2}, \,\,\hs
t_1 = \frac{k_1 (n_1-k_1 k_2^2)-1}{k_1 \left[k_1 (1 + k_2^2) n_1-1\right]}, \hs
t_2 = \frac{k_2 (k_1 + n_1)}{-1 + k_1 (1 + k_2^2) n_1} \hspace{0.2cm}\mbox{for IH,}%
\end{array}%
\right. \label{k12t2}\\
&&A=-\frac{C_2 - B_2 (k_1 + n_1) + C_1 k_1 k_2 n_2 + C_3 (k_1 k_2 + n_2) - B_1 k_1 (k_2 n_1 + n_2)}{k_1 n_1} \hspace{0.1cm}\mbox{(NH and IH),}  \hspace{0.2cm} \label{A}\\
&&B_1=\frac{C_3 + C_1 k_1 k_2}{k_1}+\frac{(C_2 - B_2 k_1 + C_3 k_1 k_2) (n_1 - t_1)}{(n_1 t_2-n_2 t_1) k_1} \hspace{0.2cm}\mbox{(NH and IH),} \label{B1}\\
&&B_2=\frac{C_2}{k_1} + C_3 k_2 \hspace{0.25cm}\mbox{(NH and IH),} \label{B2}\\
&&C_1=\frac{C_3 (k_1 - n_1)+\frac{(C_2 - B_2 n_1 + C_3 n_2) (k_1 - t_1)}{t_2-k_2 t_1}+\frac{(C_2 - B_2 k_1 + C_3 k_1 k_2)(n_1 - t_1)n_1}{n_2 t_1-n_1 t_2}}{k_1 (k_2 n_1 - n_2)} \hspace{0.1cm}\mbox{(NH and IH),} \label{C1}\\
&&C_2=\left\{
\begin{array}{l}
\frac{\sqrt{\Delta m^2_{21}}}{1 + n_1^2 + n_2^2}+\frac{\sqrt{\Delta m^2_{31}} n_2^2}{(1 + n_1 t_1)^2 + n_2^2 (1 + t_1^2)} \hspace{0.2cm}\mbox{for NH,}\ \  \\
k^2_1\left(\frac{k_2^2 \left(\sqrt{-\Delta m^2_{31}}-\sqrt{\Delta m^2_{21}-\Delta m^2_{31}}\right)}{1 + 2 k_1 n_1 + k_1^2 \big[n_1^2 + k_2^2 (1 + n_1^2)\big]}+\frac{(1 + k_2^2) \sqrt{\Delta m^2_{21}-\Delta m^2_{31}}}{1 + k_1^2 (1 + k_2^2)}\right) \hspace{0.5cm}\mbox{for IH,}%
\end{array}%
\right.\label{C2}\\
&&C_3=\left\{
\begin{array}{l}
\frac{\sqrt{\Delta m^2_{21}} n_2}{1 + n_1^2 + n_2^2}-\frac{\sqrt{\Delta m^2_{31}}(1+n_1 t_1) n_2}{(1 + n_1 t_1)^2 + n_2^2 (1 + t_1^2)} \hspace{0.2cm}\mbox{for NH,}\ \  \\
k_1 k_2\left(\frac{\left(\sqrt{\Delta m^2_{21}-\Delta m^2_{31}}-\sqrt{-\Delta m^2_{31}}\right) (k_1 n_1+1)}{1 + 2 k_1 n_1 + k_1^2 \big[n_1^2 + k_2^2 (1 + n_1^2)\big]}-\frac{\sqrt{\Delta m^2_{21}-\Delta m^2_{31}}}{1 + k_1^2 (1 + k_2^2)}\right)\hspace{0.1cm}\mbox{for IH.}%
\end{array}%
\right. \label{C3}
 \eea
The corresponding leptonic mixing matrix is
\begin{equation}
\mathrm{U}=\mathrm{U}_{L}^{\dag} \mathrm{U}_{\nu }=\left\{
\begin{array}{l}
\left(
\begin{array}{ccc}
 \frac{k_{1}}{\sqrt{\left(k_{2}^2+1\right) k_{1}^2+1}} & \frac{n_{1}}{\sqrt{n_{1}^2+n_{2}^2+1}}&\frac{t_{1}}{\sqrt{t_{1}^2+t_{2}^2+1}} \\
 \frac{c_\psi k_{1} k_{2}+e^{-i \theta} s_\psi}{\sqrt{\left(k_{2}^2+1\right) k_{1}^2+1}} & \frac{e^{-i \theta} \left(c_\psi e^{i \theta} n_{2}+s_\psi\right)}{\sqrt{n_{1}^2+n_{2}^2+1}} & \frac{e^{-i \theta} \left(s_\psi+ e^{i \theta} c_\psi t_{2}\right)}{\sqrt{t_{1}^2+t_{2}^2+1}} \\
 \frac{c_\psi-e^{i \theta} k_{1} k_{2} s_\psi}{\sqrt{\left(k_{2}^2+1\right) k_{1}^2+1}} & \frac{c_\psi-e^{i \theta} n_{2} s_\psi}{\sqrt{n_{1}^2+n_{2}^2+1}} & \frac{c_\psi-e^{i \theta} s_\psi t_{2}}{\sqrt{t_{1}^2+t_{2}^2+1}} \\
\end{array}
\right) \hspace{0.2cm}\mbox{for NH},  \label{Ulep}  \\
\left(
\begin{array}{ccc}
 \frac{n_{1}}{\sqrt{n_{1}^2+n_{2}^2+1}}&\frac{t_{1}}{\sqrt{t_{1}^2+t_{2}^2+1}}&\frac{k_{1}}{\sqrt{\left(k_{2}^2+1\right) k_{1}^2+1}}  \\
 \frac{e^{-i \theta} \left(c_\psi e^{i \theta} n_{2}+s_\psi\right)}{\sqrt{n_{1}^2+n_{2}^2+1}} & \frac{e^{-i \theta} \left(s_\psi+ e^{i \theta} c_\psi t_{2}\right)}{\sqrt{t_{1}^2+t_{2}^2+1}}&\frac{c_\psi k_{1} k_{2}+e^{-i \theta} s_\psi}{\sqrt{\left(k_{2}^2+1\right) k_{1}^2+1}} \\
  \frac{c_\psi-e^{i \theta} n_{2} s_\psi}{\sqrt{n_{1}^2+n_{2}^2+1}} & \frac{c_\psi-e^{i \theta} s_\psi t_{2}}{\sqrt{t_{1}^2+t_{2}^2+1}}&\frac{c_\psi-e^{i \theta} k_{1} k_{2} s_\psi}{\sqrt{\left(k_{2}^2+1\right) k_{1}^2+1}} \\
\end{array}
\right) \hspace{0.2cm}\mbox{for IH}.
\end{array}%
\right.
\end{equation}
The lepton mixing matrix $\mathrm{U}_{\mathrm{PMNS}}$, in the standard parametrization, take the form:
\bea
\mathrm{U}_{\mathrm{MPNS}}=\left(
\begin{array}{ccc}
 c_{13} c_{12}  &  s_{12} c_{13} & s_{13} e^{-i \delta}  \\
\hspace{-0.15 cm} -c_{23} s_{12}-e^{i \delta}  c_{12} s_{13} s_{23} & c_{12} c_{23}-e^{i \delta} s_{12} s_{13} s_{23} & c_{13} s_{23} \\
\hspace{-0.15 cm} s_{12} s_{23}-e^{i \delta}  c_{12} c_{23} s_{13} & -c_{12} s_{23}-e^{i \delta} c_{23} s_{12} s_{13} & c_{13} c_{23}  \\
\end{array}
\hspace{-0.1 cm}\right)\hspace{-0.1 cm}\left(
\begin{array}{ccc}
 1 & 0&0 \\
 0&e^{i \eta_1} & 0 \\
 0& 0 & e^{i \eta_2} \\
\end{array}
\right)\hspace{-0.1 cm}, \label{Ustandardpara}
\eea
where $s_{ij} = \sin \theta_{ij}$ and $c_{ij} = \cos \theta_{ij}$ with $\theta_{13}, \theta_{12}$ and $\theta_{23}$ are the reactor, solar and atmospheric mixing angles, respectively; $\delta_{CP}$ is the Dirac CP violation phase and $\eta_{1,2}$ are the two Majorana CP violating phases.
Comparing the entries "12" and "13" of two mixing matrices in (\ref{Ulep}) and (\ref{Ustandardpara}) we get:
\bea
&&\eta_1=0, \, \eta_2=\delta \hs \mbox{(both NH and IH)}.
\eea
The lepton mixing angles, obtained from Eqs. (\ref{Ulep}) and (\ref{Ustandardpara}), are:
\bea &&s_{13}^2=\left| \mathrm{U}_{e 3}\right|^2=\left\{
\begin{array}{l}
\frac{t_{1}^2}{t_{1}^2+t_{2}^2+1}\hspace{0.775cm}\mbox{for  NH},    \\
\frac{k^2_1}{1 + k_1^2 \big(1 + k_2^2\big)}\hspace{0.1cm}\,\mbox{for  IH},
\end{array}%
\right.  \label{s13sq}\\
&& s_{12}^2 =\frac{\left|\mathrm{U}_{e 2}\right|^2}{1-\left|\mathrm{U}_{e 3}\right|^2}=\left\{
\begin{array}{l}
\frac{n_{1}^2 \left(t_{1}^2+t_{2}^2+1\right)}{\left(t_{2}^2+1\right) \left(n_{1}^2+n_{2}^2+1\right)}\hspace{0.3cm}\mbox{for  NH},    \\
\frac{\left[1 + k_1^2 (1 + k_2^2)\right] t_1^2}{(1 + k_1^2 k_2^2) (1 + t_1^2 + t_2^2)}\hspace{0.15cm}\,\mbox{for  IH},
\end{array}%
\right. \label{s12sq}\\
&& s_{23}^2=\frac{\left| \mathrm{U}_{\mu 3}\right|^2}{1-\left| \mathrm{U}_{e 3}\right|^2}=
\left\{
\begin{array}{l}
\frac{c_\psi^2 t_{2}^2+ s_{2\psi} c_\theta t_{2} +s_\psi^2}{t_{2}^2+1}\hspace{0.925cm}\mbox{for  NH},    \\
\fr{c_\psi^2 k_1^2 k_2^2 + s_\psi^2 + k_1 k_2 s_{2\psi} c_\theta}{1 + k_1^2 k_2^2}\hspace{0.15cm}\,\mbox{for  IH},
\end{array}%
\right. \label{s23sq}\eea
The Jarlskog invariant in the active sector, 
determined from Eq. (\ref{Ulep}), takes the form \cite{PDG2022,Jarlskog1}
\bea
J^{(l)}_{CP}&=&\frac{ n_1 t_1 (t_2-n_2) s_\psi c_\psi s_\theta}{\left(n_{1}^2+n_{2}^2+1\right) \left(1 + t_1^2 + t_2^2\right)} \hspace{0.15cm}\mbox{(NH and IH).} \label{Jmex}
\eea
Comparing $J^{(l)}_{CP}$ in Eq. (\ref{Jmex}) and that of the standard parametrization, $J^{(l)}_{CP} =  c_{12} c_{13}^2 c_{23}  s_{12} s_{13} s_{23}  s_\delta$, we obtain:
\bea
&&s_\delta=\frac{ n_1 t_1 (t_2-n_2) s_\psi c_\psi s_\theta}{\left(n_{1}^2+n_{2}^2+1\right) \left(1 + t_1^2 + t_2^2\right) c_{12} c_{13}^2 c_{23} s_{12} s_{13} s_{23} } \hspace{0.15cm}\mbox{(NH and IH)}. \label{sd}
\eea
The effective neutrino masses \cite{betdecay1}, obtained from Eqs. (\ref{m1m2m3}), (\ref{Unu}) and (\ref{Ulep}), possess the following forms:
\bea
&&\langle m_{ee}\rangle = \left| \sum^3_{i=1} U_{ei}^2 m_i \right|=\left\{
\begin{array}{l}
\frac{\sqrt{\Delta m^2_{21}} n_1^2}{1 + n_1^2 + n_2^2} +\frac{\sqrt{\Delta m^2_{31}} t_1^2}{1 + t_1^2 + t_2^2} \hspace{1.4cm}\mbox{for  NH},    \\
\frac{\sqrt{-\Delta m^2_{31}} n_1^2}{1 + n_1^2 + n_2^2} + \frac{\sqrt{\Delta m^2_{21}-\Delta m^2_{31}} t_1^2}{1 + t_1^2 + t_2^2} \hspace{0.1cm}\,\mbox{for  IH},
\end{array}%
\right. \label{meeexpr}\\
&&m_{\beta }= \sqrt{\sum^3_{i=1} \left|U_{ei}\right|^2 m_i^2}=\left\{
\begin{array}{l}
\sqrt{\frac{\Delta m^2_{21} n_1^2}{1 + n_1^2 + n_2^2} +\frac{\Delta m^2_{31} t_1^2}{1 + t_1^2 + t_2^2}} \hspace{1.4cm}\mbox{for  NH},    \\
\sqrt{\frac{\left(\Delta m^2_{21}-\Delta m^2_{31}\right) t_1^2}{1 + t_1^2 + t_2^2}-\frac{\Delta m^2_{31} n_1^2}{1 + n_1^2 + n_2^2}} \hspace{0.1cm}\,\mbox{for  IH},
\end{array}%
\right. \label{mbexpr}
\eea
From Eqs. (\ref{s13sq})-(\ref{s23sq}), we can express $n_{1,2}, t_{1}$ and $s_{\delta}$ in terms of two constrained parameters $c_\theta, s_\psi$ and five observable parameters $\Delta m^2_{21}, \Delta m^2_{31}$, $s^2_{12}, s^2_{23}$, $s^2_{13}$ and  as follows:
\begin{itemize}
  \item [$\bullet$] For  NH:
  \bea
&&n_1=\frac{s_{12}^2 c_{13}^4 t_1^2}{\sqrt{\big(c_{13}^2 t_1^2-s_{13}^2\big) s_{12}^2 c_{12}^2 c_{13}^4 t_1^2 }-s_{12}^2 s_{13}^2 c_{13}^2 t_1}, \hs 
n_2=\frac{(1 + n_1 t_1) s_{13}}{\sqrt{c_{13}^2 t_1^2-s_{13}^2}}, \label{n12}\\
&&t_1=t_{13}\sqrt{\frac{s_\psi^2 (s_{23}^2 - c_\psi^2) +c_\psi^2 (c_{23}^2 + c_{2\theta} s_\psi^2)+
 2 \sqrt{c_\theta^2 c_\psi^2 s_\psi^2 (s_{23}^2 c_{23}^2 - s_\psi^2 c_\psi^2 s_\theta^2 )}}{ \big(c_\psi^2-s_{23}^2\big)^2}}. \label{t1}
\eea
\item [$\bullet$] For IH:
\end{itemize}
\bea
&&k_1=-t_{13}\sqrt{\frac{s_\psi^2 (s_{23}^2 - c_\psi^2) +c_\psi^2 (c_{23}^2 + c_{2\theta} s_\psi^2)-
 2 \sqrt{c_\theta^2 c_\psi^2 s_\psi^2 (s_{23}^2 c_{23}^2 - s_\psi^2 c_\psi^2 s_\theta^2 )}}{\left(c_{\psi}^2-s_{23}^2\right)^2}}, \label{t1}\\
&&k_2=\frac{\sqrt{k_1^2 c_{13}^2- s_{13}^2}}{k_1 s_{13}}, \hs
 n_1=\frac{s_{12} c_{12} c_{13}^2\sqrt{k_1^2\left(k_1^2 c_{13}^2-s_{13}^2\right)}-k_1 c_{12}^2 s_{13}^2 c_{13}^2}{s_{13}^4  + s_{12}^2 c_{13}^2 \left(s_{13}^2- k_1^2\right)}. \label{k2n1i}\eea
Expressions (\ref{k12t2})-(\ref{C3}) and (\ref{sd})-(\ref{k2n1i}) show that the model parameters $s_\delta, k_{1,2}, n_{1,2}$ and $t_{1,2}$ 
depend on two constrained parameters $c_\theta, s_\psi$ and three observable parameters $s^2_{12}, s^2_{23}$, $s^2_{13}$ while $A, B_{1,2}, C_{1,2, 3}, \langle m_{ee}\rangle$ and $m_{\beta}$ depend on two constrained parameters $c_\theta, s_\psi$ and five observable parameters $\Delta m^2_{21}, \Delta m^2_{31}$, $s^2_{12}, s^2_{23}$, $s^2_{13}$.
\section{\label{NR} Numerical analysis}
\emph{$\bullet$ For the charged lepton sector}, using the values of $\Lambda$, the observed values of the charged lepton masses \cite{PDG2022}, $m_e=0.51099 \,\mathrm{MeV},  m_\mu = 105.65837\,\mathrm{MeV}, m_\tau = 1776.86 \,\mathrm{MeV}$ and the VEV of scalar fields in Eqs. (\ref{vev2H}) and (\ref{vevscales}), with the help of Eqs. (\ref{aiexpres}) and (\ref{a0b0c0d0})-(\ref{BCDG0expres}), we get $|h_1|\simeq 10^{-2}$, and $h_{2,3,4,5}$ are still depend on three parameters $\alpha$, $\theta$ and $\psi$. In the case of $s_\alpha=-0.95\, (\alpha=288.2^\circ)$, the Yukawa-like couplings $h_{2,3,4,5}$ depend on two parameters $\theta$ and $\psi$ which are plotted in Figs. \ref{h2h3f} and \ref{h4h5f}.
\begin{center}
\begin{figure}[h]
\vspace{-0.75 cm}
\hspace{-5.5 cm}
\includegraphics[width=0.8\textwidth]{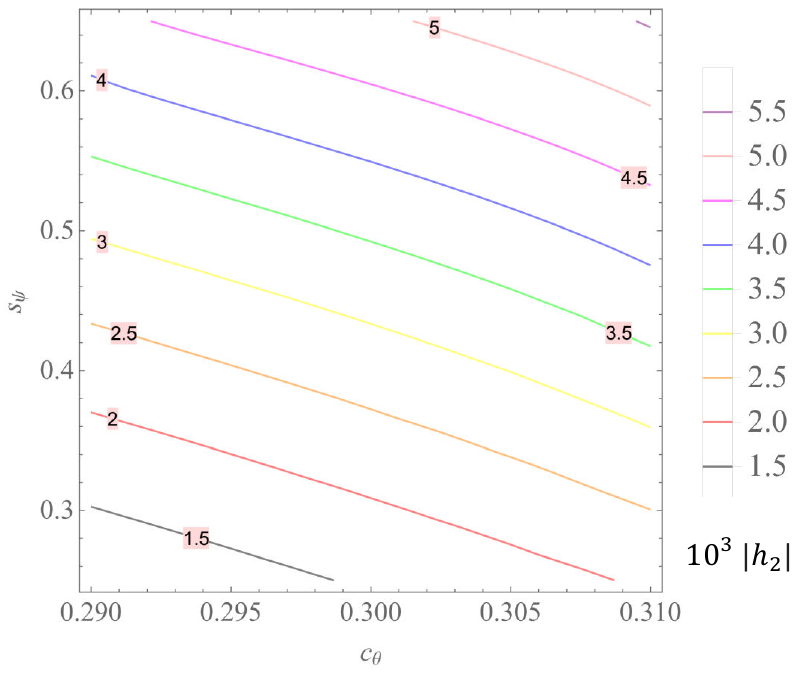}\hspace{-4.85 cm}
\includegraphics[width=0.8\textwidth]{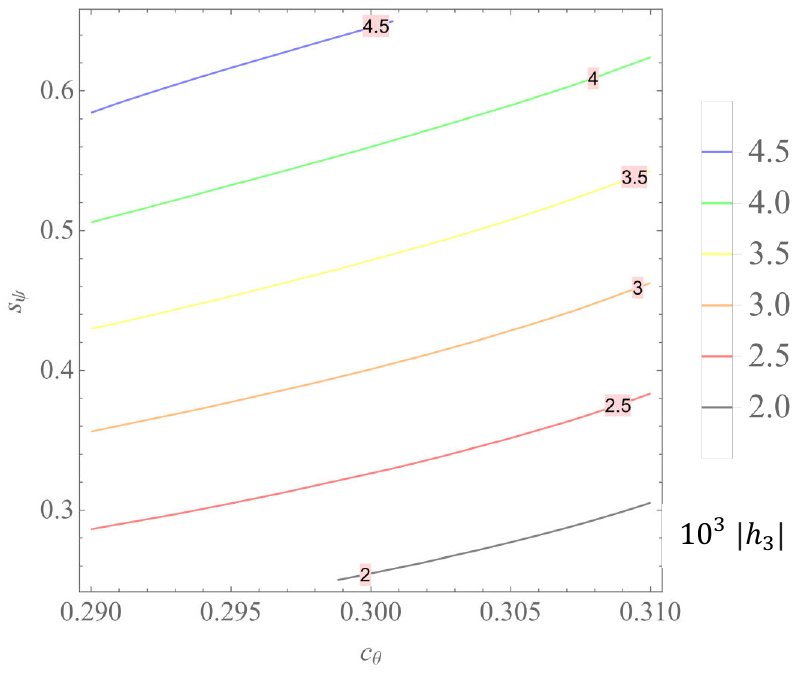}\hspace{-6.25 cm}
\vspace{-9.25 cm}
\caption{$10^3|h_2|$ (left panel) and $10^3|h_3|$ (right panel) versus $c_\theta$ and $s_{\psi}$ with $c_\theta\in(0.29, 0.31)$ and $s_\psi\in (0.25, 0.65)$.}
\label{h2h3f}
\end{figure}
\vspace{-0.5 cm}
\end{center}
\begin{center}
\begin{figure}[h]
\vspace{-1.25 cm}
\hspace{-5.5 cm}
\includegraphics[width=0.8\textwidth]{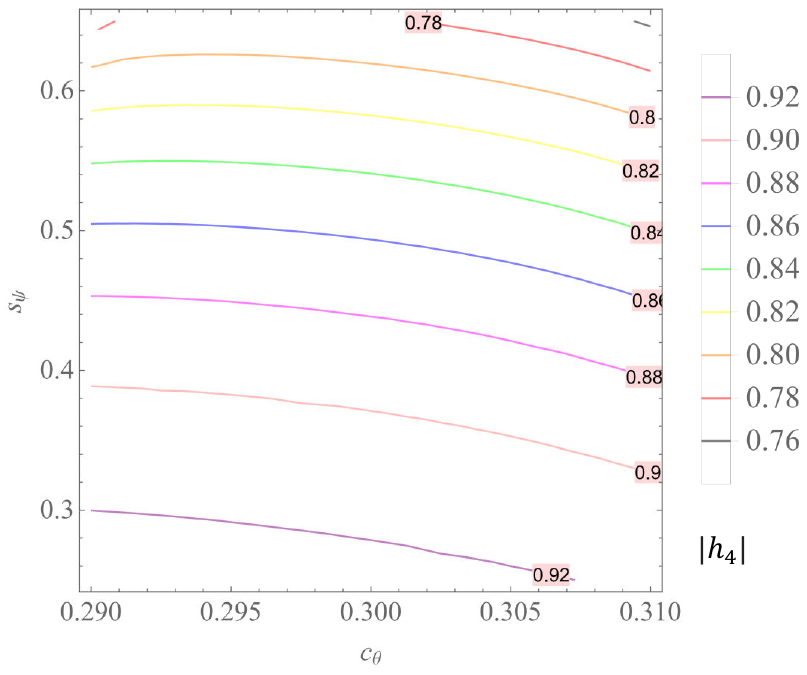}\hspace{-4.8 cm}
\includegraphics[width=0.8\textwidth]{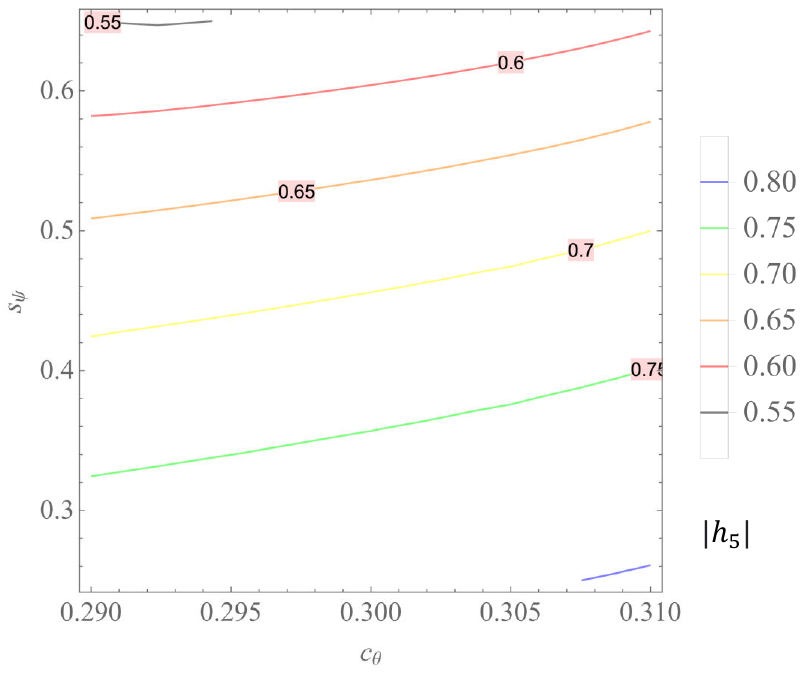}\hspace{-6.25 cm}
\vspace{-9.25 cm}
\caption{$|h_4|$ (left panel) and $|h_5|$ (right panel) versus $c_\theta$ and $s_{\psi}$ with $c_\theta\in(0.29, 0.31)$ and $s_\psi\in (0.25, 0.65)$.}
\label{h4h5f}
\end{figure}
\vspace{-0.5 cm}
\end{center}
Figures \ref{h2h3f} and \ref{h4h5f} imply
\bea
&&|h_2|\simeq |h_3| \sim 10^{-2}, \hs |h_4|\simeq |h_5| \sim 10^{-1}, \label{h2h3h4ranges}
\eea
which implies that the Yukawa couplings in the charged lepton sector differ from each other by one order of magnitude for a natural explanation to the charged lepton mass hierarchy.

\emph{$\bullet$ For neutrino sector}. Equation (\ref{m1m2m3}) shows that neutrino masses
($m_{2,3}$ for NH and $m_{1,2}$ for IH) depend on two experimental parameters $\Delta m^2_{31}$ and $\Delta m^2_{21}$ which have been measured with high accuracy.
In the case of $\Delta m^2_{21}$ and $\Delta m^2_{31}$ lie in 3$\sigma$ range \cite{Salas2021}, i.e., $\Delta m^2_{21}\in (69.40, 81.40)\, \mathrm{meV^2}$  and $\Delta m^2_{31}\in (2.47, 3.63)10^3\, \, \mathrm{meV^2}$, we get the allowed
regions for $m_{1,2,3}$, $m_1=0,\hs m_2\in  (8.33, 9.02)\, \mathrm{meV}$, $m_3= (49.70, 51.30) \, \mathrm{meV}$ for  NH, and $m_1\in (48.70, 50.30)\, \mathrm{meV},\,  m_2= (49.4,51.0)\, \mathrm{meV}, \hs m_3=0$ for  IH. The sum of neutrino masses are predicted to be
\bea
&&\sum m_{\nu} \, (\mathrm{meV}) \in \left\{
\begin{array}{l}
(58.25, 60.25) \hspace{0.2cm}\mbox{for  NH},    \\
(98.50, 101.0) \hspace{0.2cm}\mbox{for  IH},
\end{array}%
\right. \label{sumrange}
\eea
which are in consistent with the limits \cite{RoyChoudhury} 
$\sum m_{\nu}< 0.15$ eV (NH) and $\sum m_{\nu}< 0.17$ eV (IH),
$\sum m_{\nu} < 0.14 $ eV \cite{Tanseri22}, $\sum m_{\nu} < 0.152$ eV \cite{nubound} (minimal $\Lambda \mathrm{CDM} + \sum m_{\nu}$), $\sum m_{\nu} < 0.118$ eV (high-$l$ polarization), $\sum m_{\nu} < 0.101$ eV (NPDDE model), $\sum m_{\nu} <$ 0.093 eV (NPDDE+$r$ model) and the most aggressive bound is $\sum m_{\nu} < 0.078 \mathrm{eV}$ (NPDDE+r with the R16 prior) \cite{nubound, Vagnozzisum}, $\sum m_{\nu} < 0.183$ eV for IH \cite{Giusarmasum16}, $\sum m_{\nu} < 0.13$ eV (the base dataset) and $\sum m_{\nu} <
0.11$ eV (pol dataset) \cite{Vagnozzisum18}, $\sum m_{\nu} < 0.19$ eV \cite{Giusarmasum18}.

In order to determine the possible ranges of the parameters $k_{1,2}, n_{1,2}, t_{1,2}$ and get predictive values for the Dirac CP viloation phase $\delta$, we use the observables $\Delta m^2_{21}$, $\Delta m^2_{31}$, $\sin^2 \theta_{12}$, $\sin^2\theta_{23}$ and $\sin^2 \theta_{13}$, whose experimental values given in Table \ref{Salas2021T}, as input parameters.
\begin{table}[ht]
\vspace{-0.25cm}
\begin{center}
\caption{\label{Salas2021T} The global analysis of neutrino oscillation data \cite{Salas2021} }
\vspace{0.25cm}
\begin{tabular}{|c|c|c|c|c|c|c|c|c|c|c|}\hline
&$\mathrm{Best-fit \,\,point}\, (3\sigma \,\, \mathrm{range}) \,(\mathrm{NH})$  & $\mathrm{Best-fit \,\, point}\, (3\sigma \,\, \mathrm{range}) \,(\mathrm{IH})$ \\  \hline
 $\Delta m^2_{21} \left[\mathrm{meV}^2\right]$&$75.0\, (69.4\rightarrow  81.4)$&$75.0\, (69.4\rightarrow 81.4)$ \\
$\fr{|\Delta m^2_{31}| \,\left[\mathrm{meV}^2\right]}{10^{3}}$&$2.55 \, (2.47\rightarrow  2.63)$& $2.45 \, (2.37 \rightarrow 2.53)$ \\ \hline
$\sin^2\theta_{12}$&\hspace{0.1cm}$0.318\, (0.271\rightarrow 0.369)$ \hspace{0.1cm}&\hspace{0.1cm} $0.318\, (0.271\rightarrow 0.369)$ \hspace{0.1cm}\\
$\sin^2\theta_{23}$&\hspace{0.1cm}$0.574 \, (0.434\rightarrow 0.610)$ \hspace{0.1cm}&\hspace{0.1cm} $0.578\, (0.433\rightarrow 0.608)$ \hspace{0.1cm}\\
$\fr{\sin^2\theta_{13}}{10^{-2}}$&  $2.200\, (2.00\rightarrow  2.405)$& $2.225\, (2.018\rightarrow 2.424)$\\
$\delta_{CP}/\pi$&  $1.08\, (0.71\rightarrow  1.99)$ & $1.58\, (1.11\rightarrow  1.96)$ \\  \hline
\end{tabular}
\end{center}
\vspace{-0.25cm}
\end{table}

At the best-fit values of the lepton mixing angles\cite{Salas2021}, $\sin^2\theta_{12}=0.318$ and $\sin^2\theta_{13}=2.200\times 10^{-2}$ for NH while $\sin^2\theta_{12}=0.318$ and $\sin^2\theta_{13}=2.225\times 10^{-2}$ for IH, $s_\delta, k_{1,2}, n_{1,2}$ and $t_{1,2}$ depend on two parameters $c_\theta$ and $s_\psi$. The Dirac CP violating phase $\delta$ (more precisely, $s_\delta$) as a function of two parameters $c_{\theta}$ and $s_{\psi}$, with $c_\theta\in(0.29, 0.31)$ and $s_\psi\in (0.25, 0.65)$ for both IH and  NH, is plotted in Fig. \ref{sdf}, which implies that
\bea
&&s_\delta\in  (-0.95, \, -0.50),\,\, \mathrm{i.e.},\,\, \delta^\circ\in  (288.20, \, 330.00)\, \hspace{0.250cm}\mbox{(NH and IH)}.
\eea
\begin{center}
\begin{figure}[h]
\vspace{-0.25 cm}
\hspace{-5.65 cm}
\includegraphics[width=0.75\textwidth]{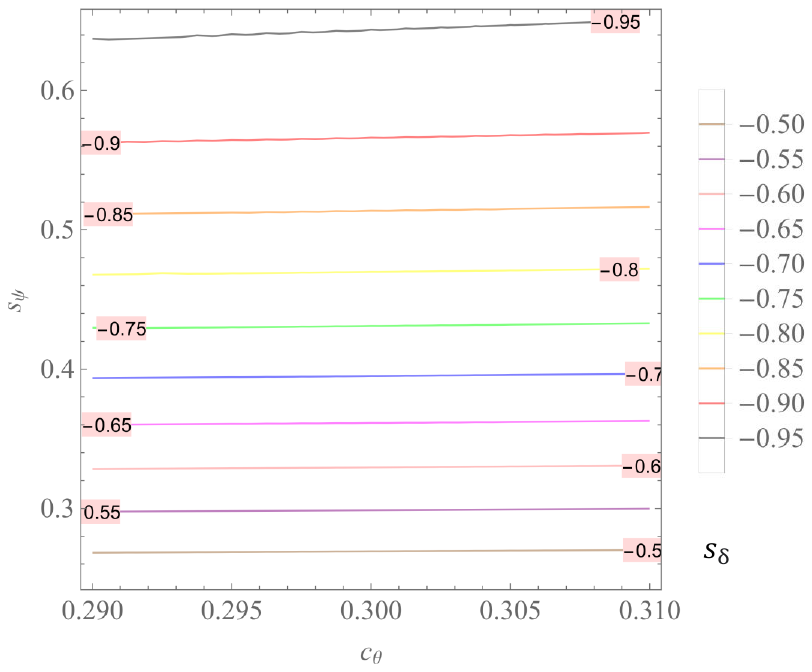}\hspace{-5.65 cm}
\vspace{-8.75 cm}
\caption{$s_\delta$ versus $c_{\theta}$ and $s_{\psi}$ with $c_\theta\in(0.29, 0.31)$ and $s_\psi\in (0.25, 0.65)$ for both NH and IH.}
\label{sdf}
\end{figure}
\vspace{-0.5 cm}
\end{center}
The dependence of $k_{1,2}, n_{1,2}$ and $t_{1,2}$ on two parameters $c_{\theta}$ and $s_{\psi}$, with $c_\theta\in(0.29, 0.31)$ and $s_\psi\in (0.25, 0.65)$ for both IH and  NH, are respectively plotted in Figs. \ref{k1f},\ref{k2f}, \ref{n1f},\ref{n2f}, \ref{t1f} and \ref{t2f}. 
\newpage
\begin{center}
\begin{figure}[h]
\vspace{-1.25 cm}
\hspace{-5.65 cm}
\includegraphics[width=0.75\textwidth]{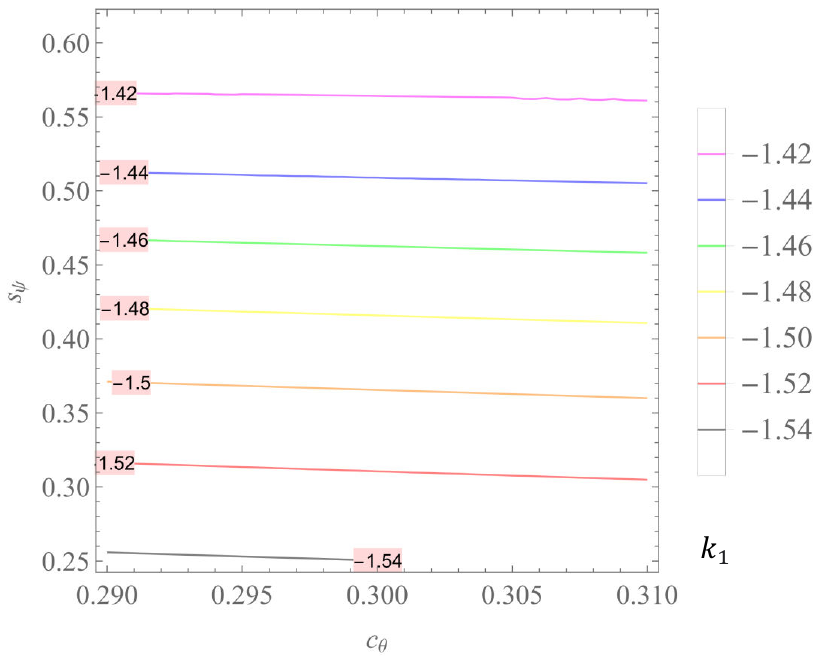}\hspace{-4.35 cm}
\includegraphics[width=0.75\textwidth]{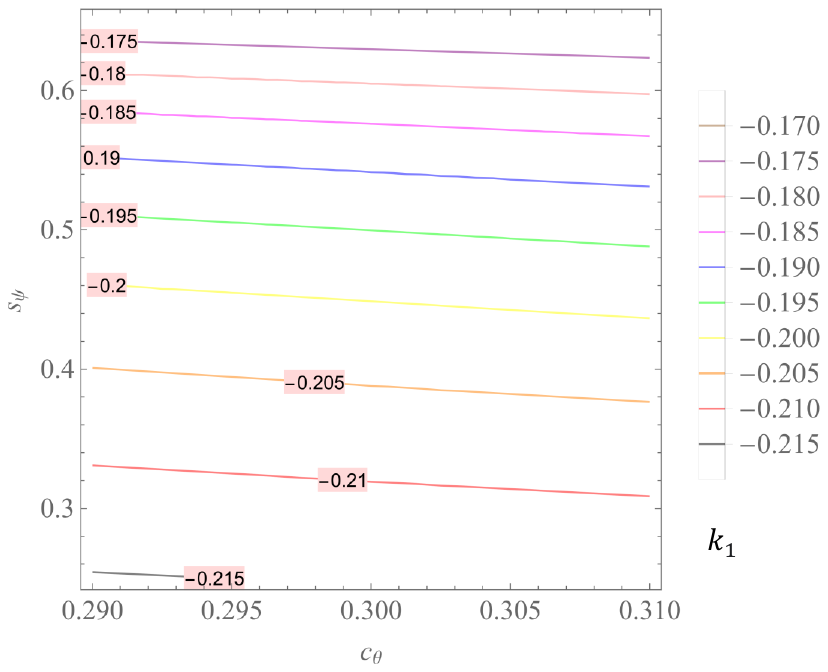}\hspace{-6.25 cm}
\vspace{-8.75 cm}
\caption{$k_1$ versus $c_{\theta}$ and $s_{\psi}$ with $c_\theta\in(0.29, 0.31)$ and $s_\psi\in (0.25, 0.65)$ for NH (left panel) and  IH (right panel).}
\label{k1f}
\end{figure}
\vspace{-1.25 cm}
\end{center}
\begin{center}
\begin{figure}[h]
\vspace{-1.0 cm}
\hspace{-5.65 cm}
\includegraphics[width=0.75\textwidth]{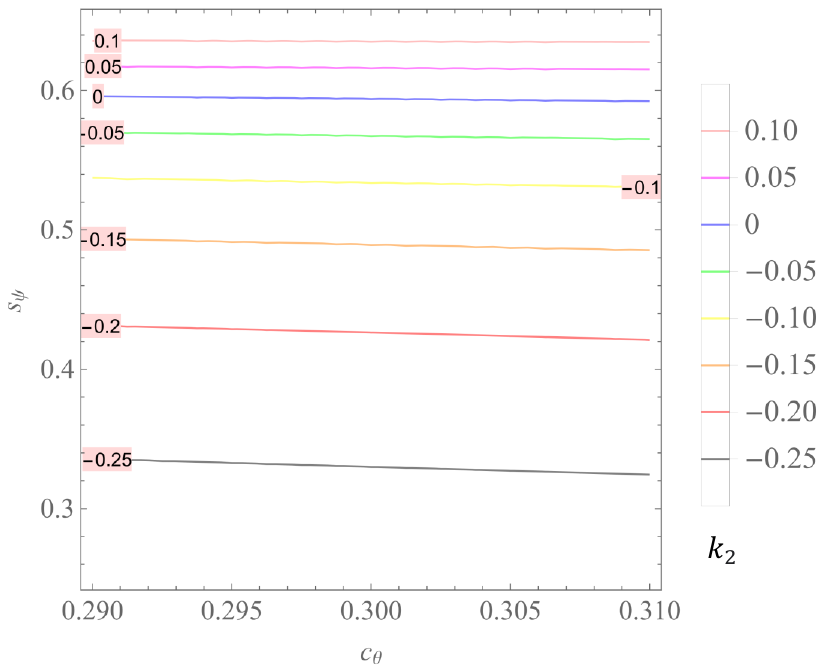}\hspace{-4.35 cm}
\includegraphics[width=0.75\textwidth]{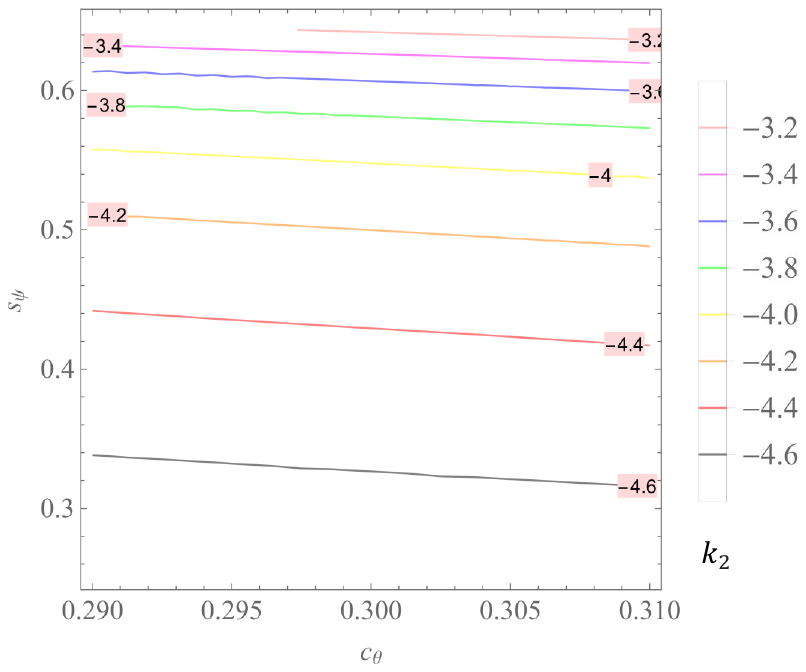}\hspace{-6.25 cm}
\vspace{-8.75 cm}
\caption{$k_2$ versus $c_{\theta}$ and $s_{\psi}$ with $c_\theta\in(0.29, 0.31)$ and $s_\psi\in (0.25, 0.65)$ for NH (left panel) and IH (right panel).}
\label{k2f}
\end{figure}
\vspace{-1.5 cm}
\end{center}
\begin{center}
\begin{figure}[h]
\vspace{-1.0 cm}
\hspace{-5.65 cm}
\includegraphics[width=0.75\textwidth]{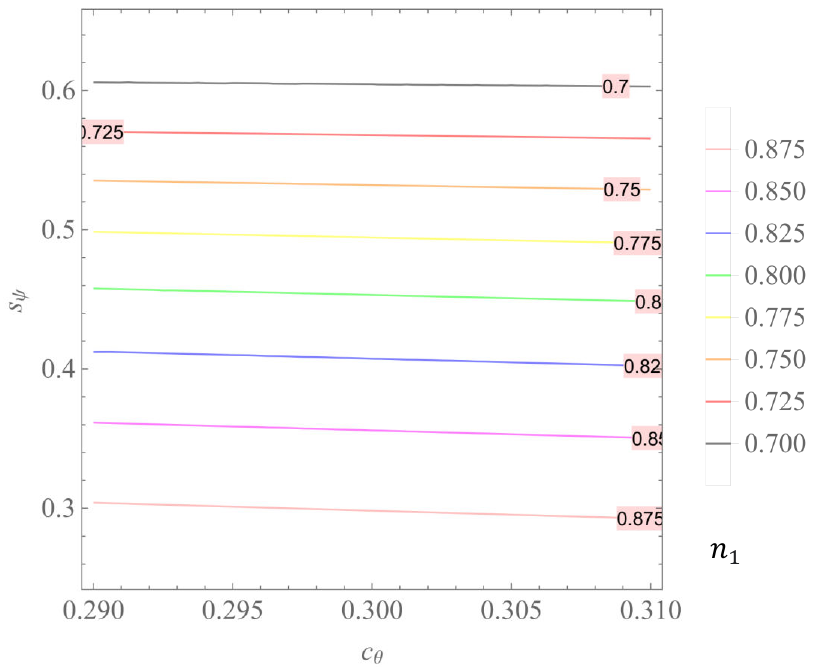}\hspace{-4.35 cm}
\includegraphics[width=0.75\textwidth]{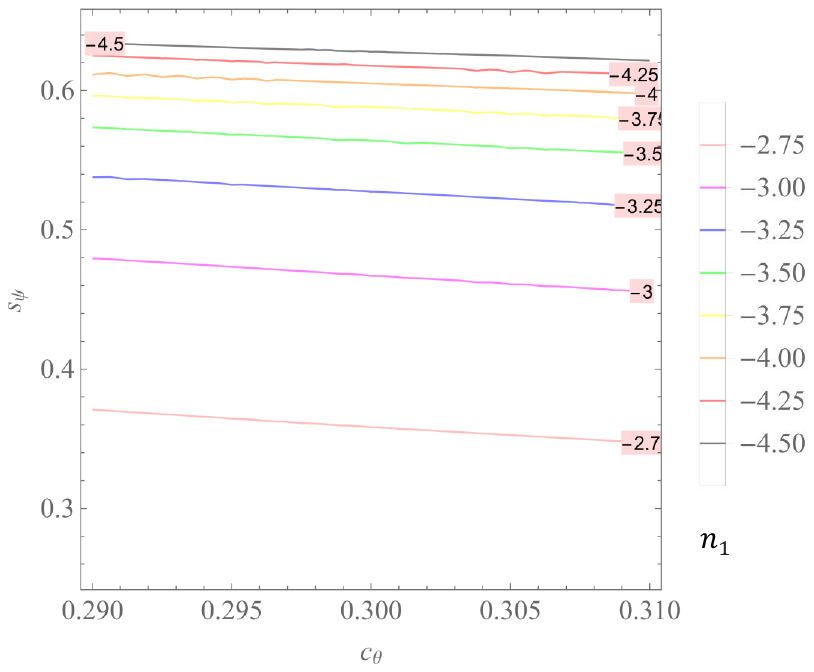}\hspace{-6.25 cm}
\vspace{-9.0 cm}
\caption{$n_1$ versus $c_{\theta}$ and $s_{\psi}$ with $c_\theta\in(0.29, 0.31)$ and $s_\psi\in (0.25, 0.65)$ for NH (left panel) and IH (right panel).}
\label{n1f}
\end{figure}
\vspace{-0.5 cm}
\end{center}
\newpage
\begin{center}
\begin{figure}[h]
\vspace{-1.0 cm}
\hspace{-5.65 cm}
\includegraphics[width=0.75\textwidth]{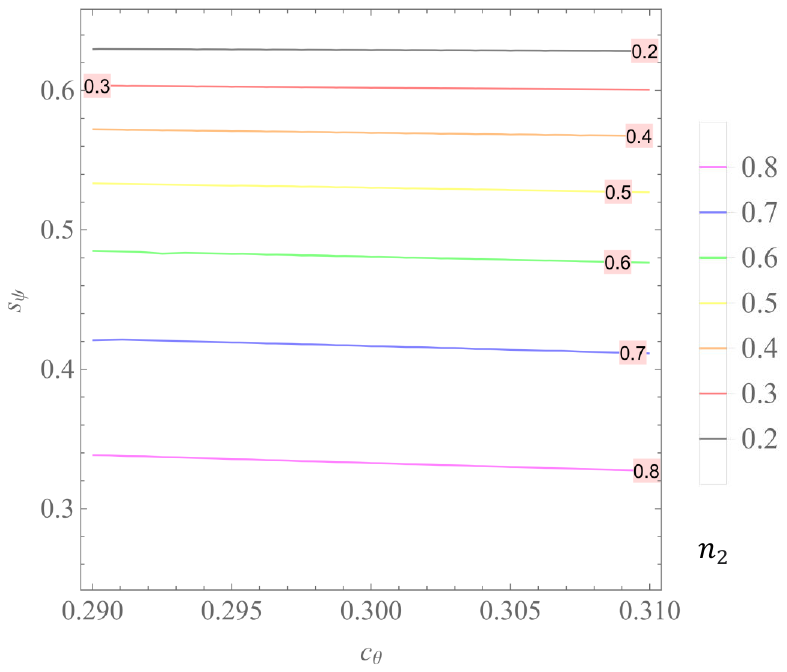}\hspace{-4.35 cm}
\includegraphics[width=0.75\textwidth]{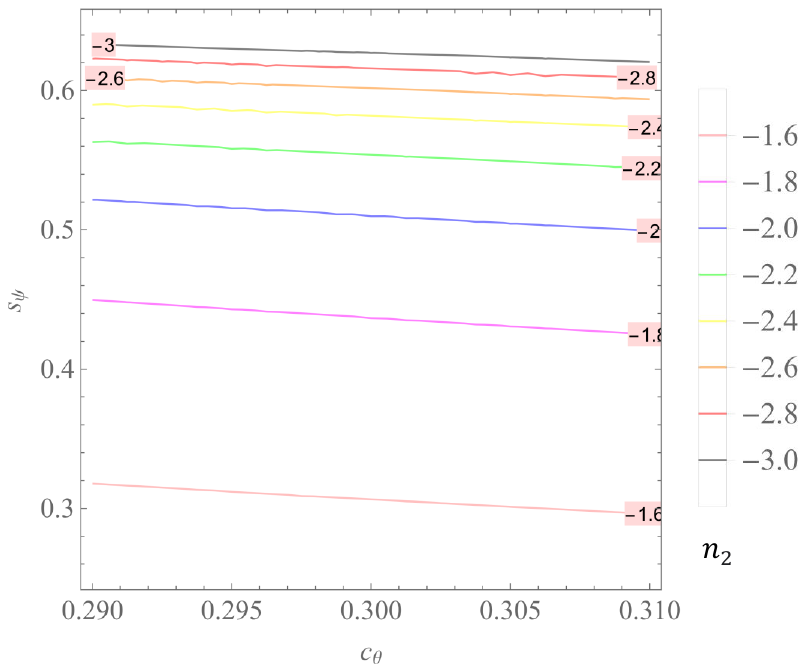}\hspace{-6.25 cm}
\vspace{-9.0 cm}
\caption{$n_2$ versus $c_{\theta}$ and $s_{\psi}$ with $c_\theta\in(0.29, 0.31)$ and $s_\psi\in (0.25, 0.65)$ for NH (left panel) and IH (right panel).}
\label{n2f}
\end{figure}
\vspace{-0.5 cm}
\end{center}
\begin{center}
\begin{figure}[h]
\vspace{-1.0 cm}
\hspace{-5.65 cm}
\includegraphics[width=0.75\textwidth]{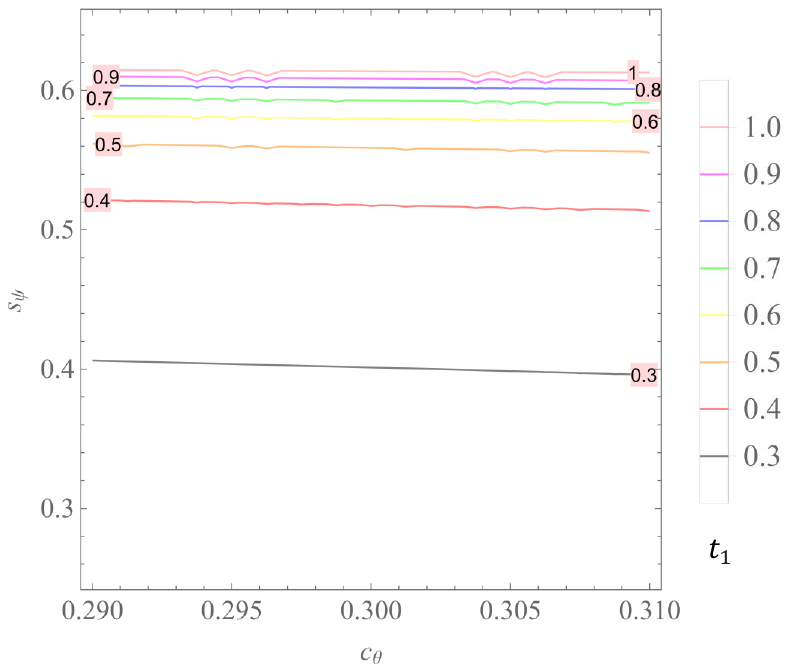}\hspace{-4.6 cm}
\includegraphics[width=0.75\textwidth]{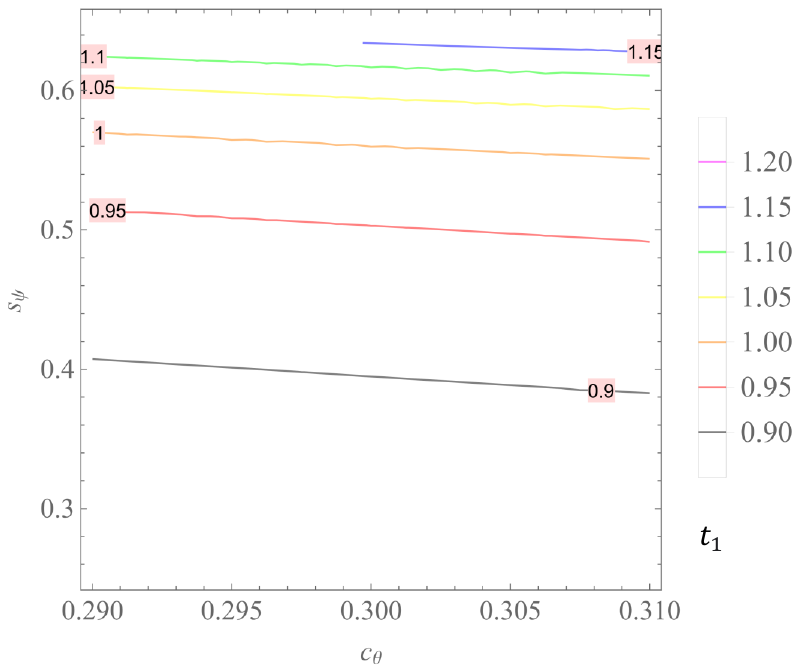}\hspace{-6.25 cm}
\vspace{-9.0 cm}
\caption{$t_1$ versus $c_{\theta}$ and $s_{\psi}$ with $c_\theta\in(0.29, 0.31)$ and $s_\psi\in (0.25, 0.65)$ for NH (left panel) and IH (right panel).}
\label{t1f}
\end{figure}
\vspace{-0.5 cm}
\end{center}
\begin{center}
\begin{figure}[h]
\vspace{-1.0 cm}
\hspace{-5.65 cm}
\includegraphics[width=0.75\textwidth]{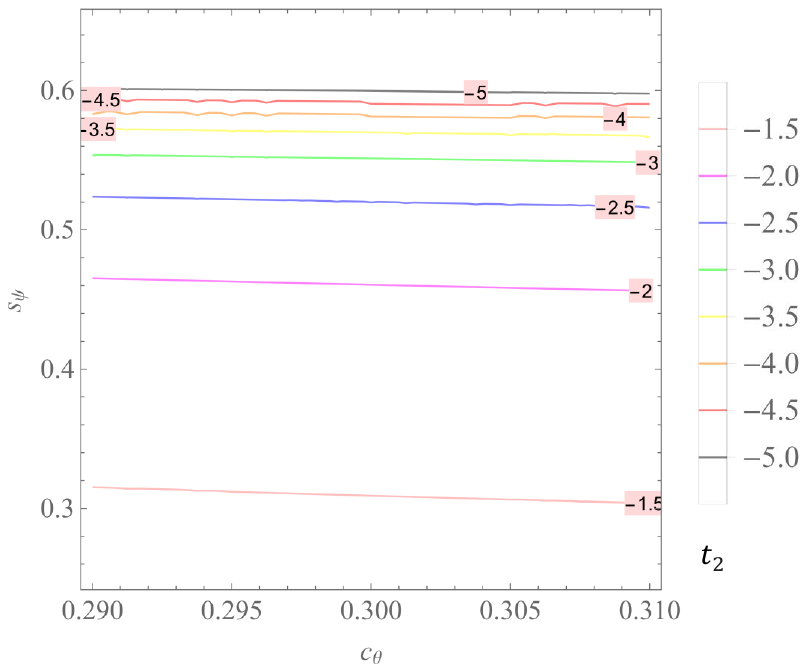}\hspace{-4.35 cm}
\includegraphics[width=0.75\textwidth]{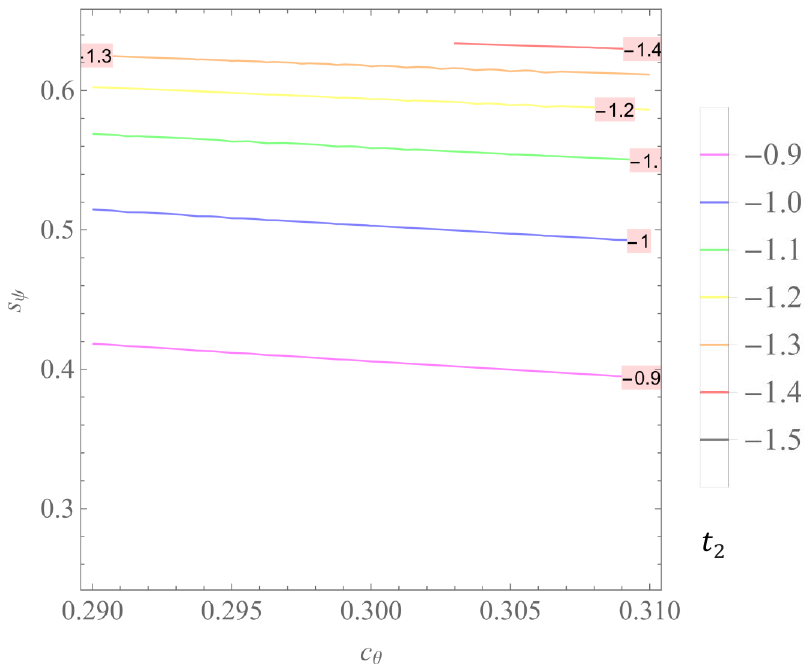}\hspace{-6.25 cm}
\vspace{-9.0 cm}
\caption{$t_2$ versus $c_{\theta}$ and $s_{\psi}$ with $c_\theta\in(0.29, 0.31)$ and $s_\psi\in (0.25, 0.65)$ for NH (left panel) and IH (right panel).}
\label{t2f}
\end{figure}
\vspace{-0.5 cm}
\end{center}
These figures imply:
\bea
&&k_{1}\in \left\{
\begin{array}{l}
(-1.54, -1.42) \hspace{0.55cm}\mbox{for  NH},  \\
(-0.215, -0.170)\hspace{0.15cm}\mbox{for IH},
\end{array}%
\right. \hspace{0.15 cm} k_{2}\in \left\{
\begin{array}{l}
(-0.25, 0.10) \hspace{0.45cm}\mbox{for  NH},  \\
(-4.60, -3.20)\hspace{0.15cm}\mbox{for IH},
\end{array}%
\right. \\
&&n_{1}\in \left\{
\begin{array}{l}
(0.70, 0.875) \hspace{0.55cm}\mbox{for  NH},  \\
(-4.50, -2.75)\hspace{0.15cm}\mbox{for IH},
\end{array}%
\right. \hspace{0.5 cm} n_{2}\in \left\{
\begin{array}{l}
(0.20, 0.80) \hspace{0.75cm}\mbox{for  NH},  \\
(-3.00, -1.60)\hspace{0.15cm}\mbox{for IH},
\end{array}%
\right.\\
&&t_{1}\in \left\{
\begin{array}{l}
(0.30,1.00) \hspace{0.40cm}\mbox{for  NH},  \\
(0.90, 1.20)\hspace{0.4cm}\mbox{for IH},
\end{array}%
\right. \hspace{0.95 cm} t_{2}\in \left\{
\begin{array}{l}
(-5.00,-1.50) \hspace{0.25cm}\mbox{for  NH},  \\
(-1.50, -0.90)\hspace{0.25cm}\mbox{for IH}.
\end{array}%
\right.\label{KKN12values}
\eea
Similarly, to determine the possible ranges of the parameters $A, B_{1,2}, C_{1,2, 3}, \langle m_{ee}\rangle$ and $m_{\beta}$ we fix $\sin^2 \theta_{12}$, $\sin^2\theta_{23}$ and $\sin^2 \theta_{13}$ at their best-fit points \cite{Salas2021} and $c_\theta=0.30\, (\theta=72.54^\circ)$ and $s_\psi=0.40\, (\psi=23.58^\circ)$ for both IH and NH, and $\Delta m^2_{21}$ and $\Delta m^2_{31}$ take the values in their 3 $\sigma$ ranges \cite{Salas2021}, $\Delta m^2_{21}\in (69.4, 81.4)\, \mathrm{meV}^2$ and $\Delta m^2_{31}\in (2.47, 2.63) 10^3\, \mathrm{meV}^2$ (NH) while $\Delta m^2_{31}\in (-2.53, -2.37) 10^3\, \mathrm{meV}^2$ (IH). The dependence of $A, B_{1,2}, C_{1,2, 3}, \langle m_{ee}\rangle$ and $m_{\beta}$ on two parameters $\Delta m^2_{21}$ and $\Delta m^2_{31}$ are presented in Figs. \ref{af}, \ref{b1f},\ref{b2f}, \ref{c1f},\ref{c2f}, \ref{c3f}, \ref{meef} and \ref{mbf}, respectively.
\begin{center}
\begin{figure}[h]
\vspace{-0.5 cm}
\hspace{-5.65 cm}
\includegraphics[width=0.75\textwidth]{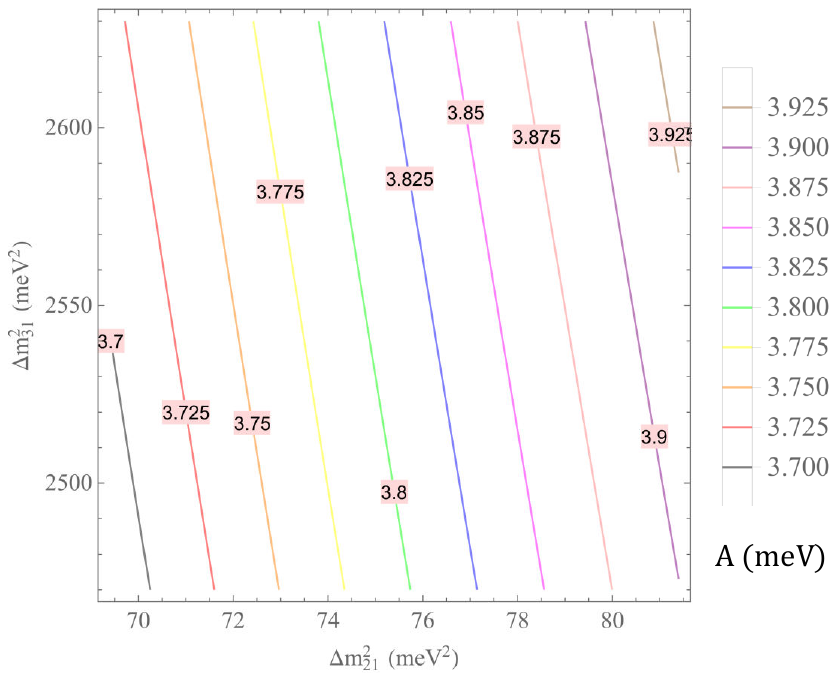}\hspace{-4.25 cm}
\includegraphics[width=0.75\textwidth]{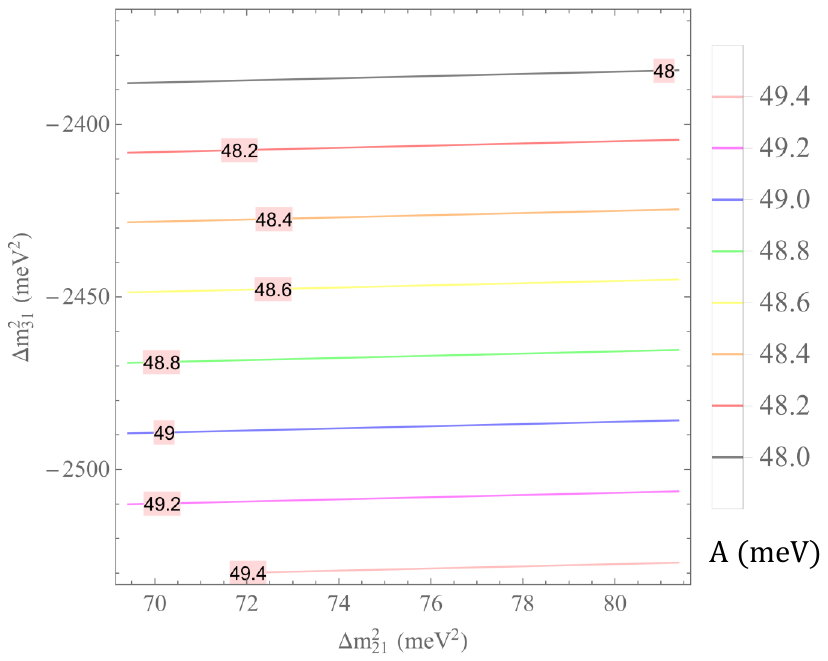}\hspace{-6.25 cm}
\vspace{-8.75 cm}
\caption{$A$\,($\mathrm{meV}$) versus $\Delta m^2_{21}$ and $\Delta m^2_{31}$ with $\Delta m^2_{21}\in (69.4, 81.4)\, \mathrm{meV}^2$ and $\Delta m^2_{31}\in (2.47, 2.63) 10^3\, \mathrm{meV}^2$ for NH (left panel) and $\Delta m^2_{31}\in (-2.53, -2.37) 10^3\, \mathrm{meV}^2$ for IH (right panel).}
\label{af}
\end{figure}
\vspace{-0.5 cm}
\end{center}
\begin{center}
\begin{figure}[h]
\vspace{-0.75 cm}
\hspace{-5.65 cm}
\includegraphics[width=0.75\textwidth]{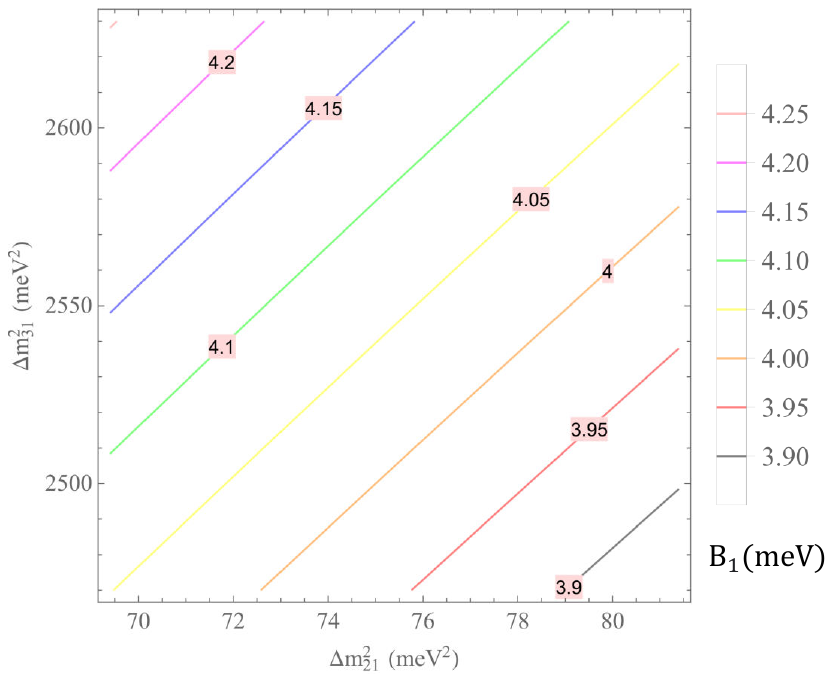}\hspace{-4.35 cm}
\includegraphics[width=0.75\textwidth]{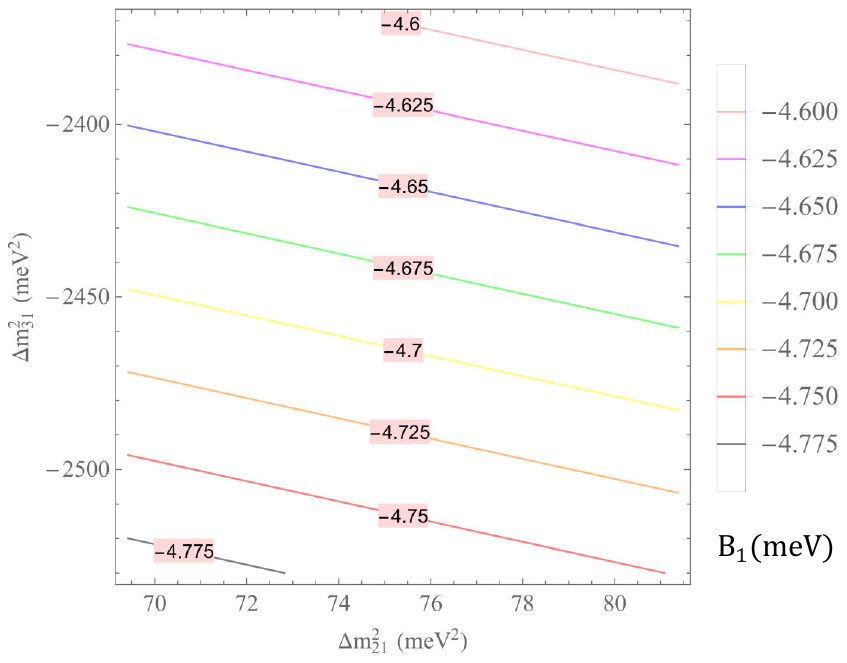}\hspace{-6.25 cm}
\vspace{-8.75 cm}
\caption{$B_1$\,($\mathrm{meV}$) versus $\Delta m^2_{21}$ and $\Delta m^2_{31}$ with $\Delta m^2_{21}\in (69.4, 81.4)\, \mathrm{meV}^2$ and $\Delta m^2_{31}\in (2.47, 2.63) 10^3\, \mathrm{meV}^2$ for NH (left panel) and $\Delta m^2_{31}\in (-2.53, -2.37) 10^3\, \mathrm{meV}^2$ for IH (right panel).}
\label{b1f}
\end{figure}
\vspace{-0.5 cm}
\end{center}
\begin{center}
\begin{figure}[h]
\vspace{-0.5 cm}
\hspace{-5.65 cm}
\includegraphics[width=0.75\textwidth]{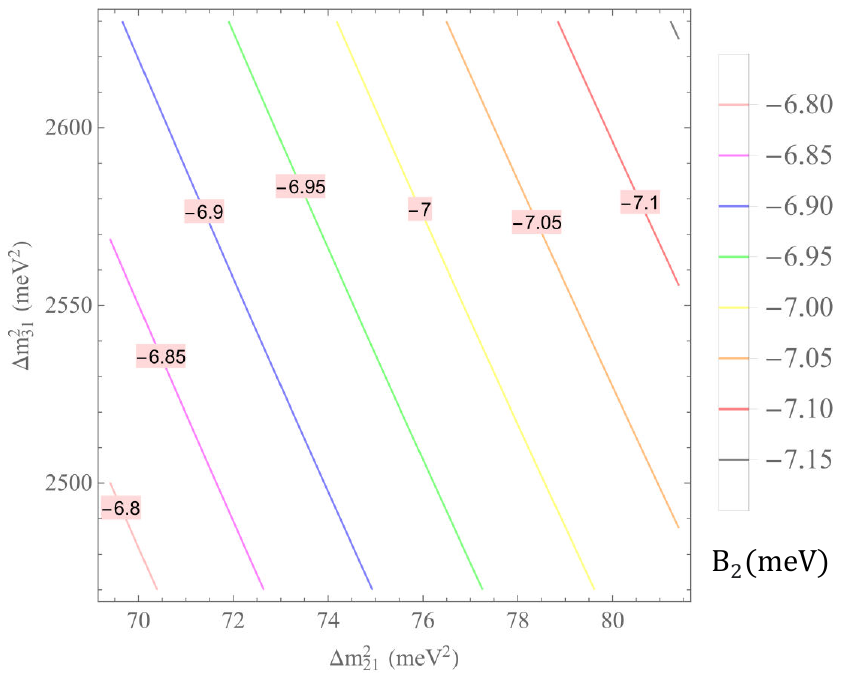}\hspace{-4.25 cm}
\includegraphics[width=0.75\textwidth]{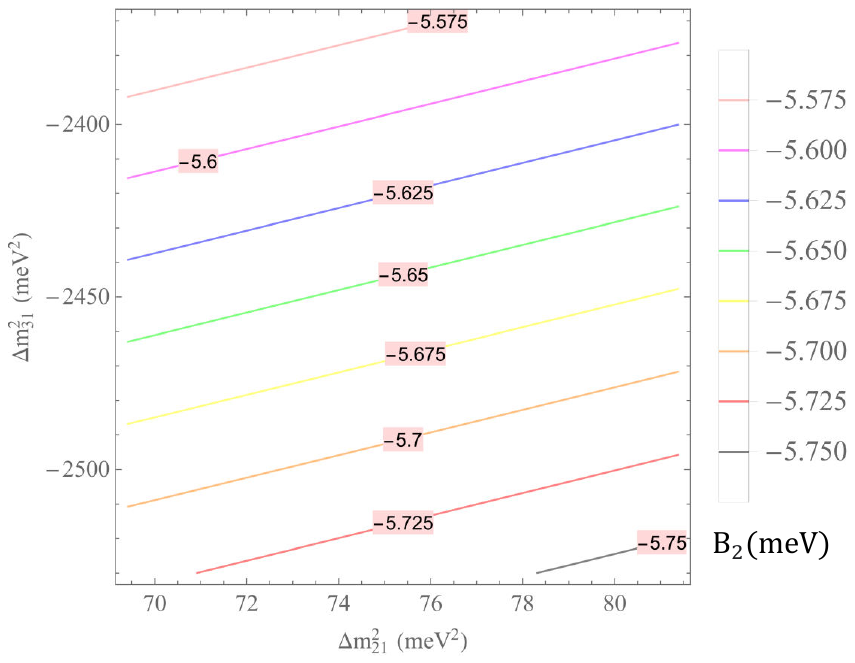}\hspace{-6.25 cm}
\vspace{-8.75 cm}
\caption{$B_2$\,($\mathrm{meV}$) versus $\Delta m^2_{21}$ and $\Delta m^2_{31}$ with $\Delta m^2_{21}\in (69.4, 81.4)\, \mathrm{meV}^2$ and $\Delta m^2_{31}\in (2.47, 2.63) 10^3\, \mathrm{meV}^2$ for NH (left panel) and $\Delta m^2_{31}\in (-2.53, -2.37) 10^3\, \mathrm{meV}^2$ for IH (right panel).}
\label{b2f}
\end{figure}
\vspace{-0.5 cm}
\end{center}
\begin{center}
\begin{figure}[h]
\vspace{-0.5 cm}
\hspace{-5.65 cm}
\includegraphics[width=0.75\textwidth]{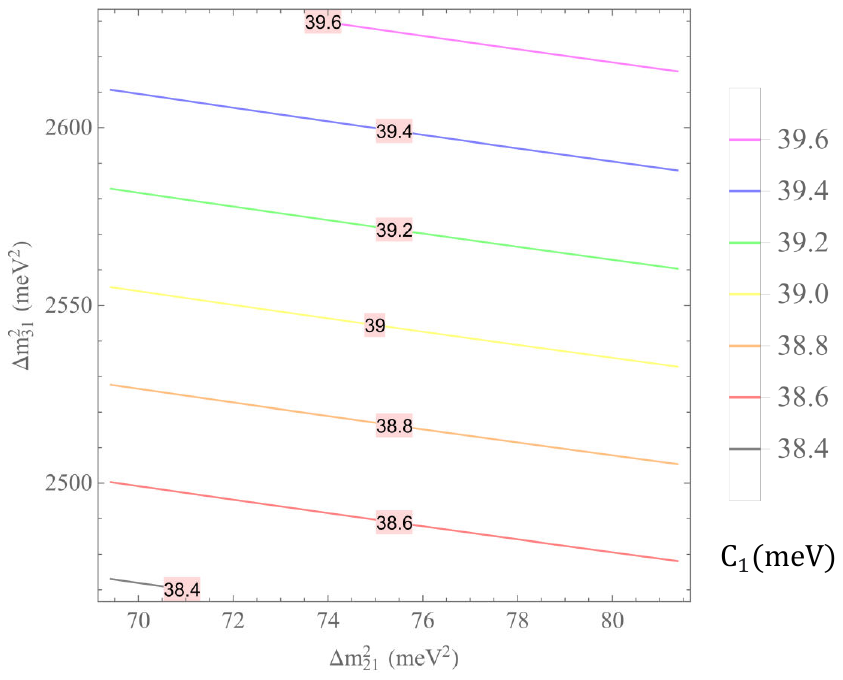}\hspace{-4.15 cm}
\includegraphics[width=0.75\textwidth]{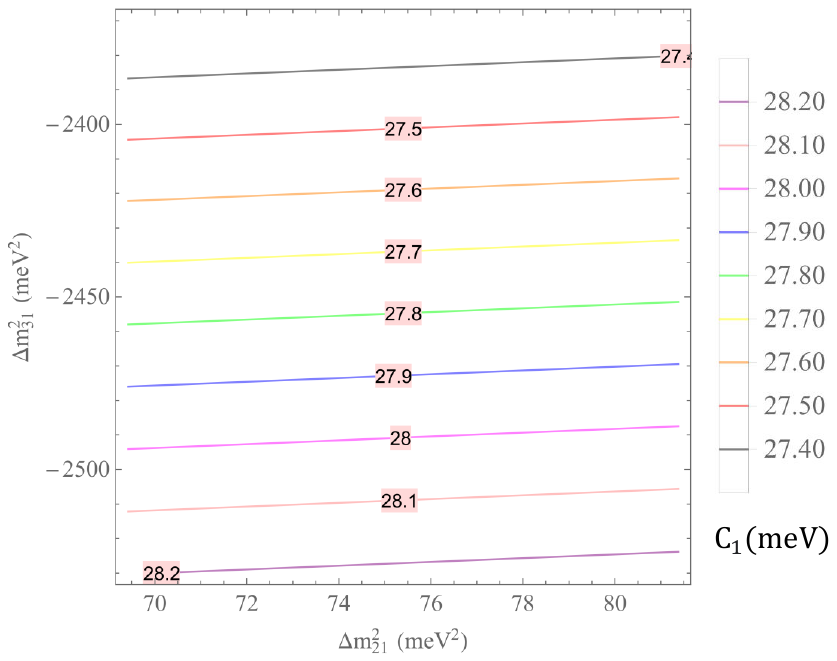}\hspace{-6.25 cm}
\vspace{-8.75 cm}
\caption{$C_1$\,($\mathrm{meV}$) versus $\Delta m^2_{21}$ and $\Delta m^2_{31}$ with $\Delta m^2_{21}\in (69.4, 81.4)\, \mathrm{meV}^2$ and $\Delta m^2_{31}\in (2.47, 2.63) 10^3\, \mathrm{meV}^2$ for NH (left panel) and $\Delta m^2_{31}\in (-2.53, -2.37) 10^3\, \mathrm{meV}^2$ for IH (right panel).}
\label{c1f}
\end{figure}
\vspace{-0.5 cm}
\end{center}
\begin{center}
\begin{figure}[h]
\vspace{-0.5 cm}
\hspace{-5.65 cm}
\includegraphics[width=0.75\textwidth]{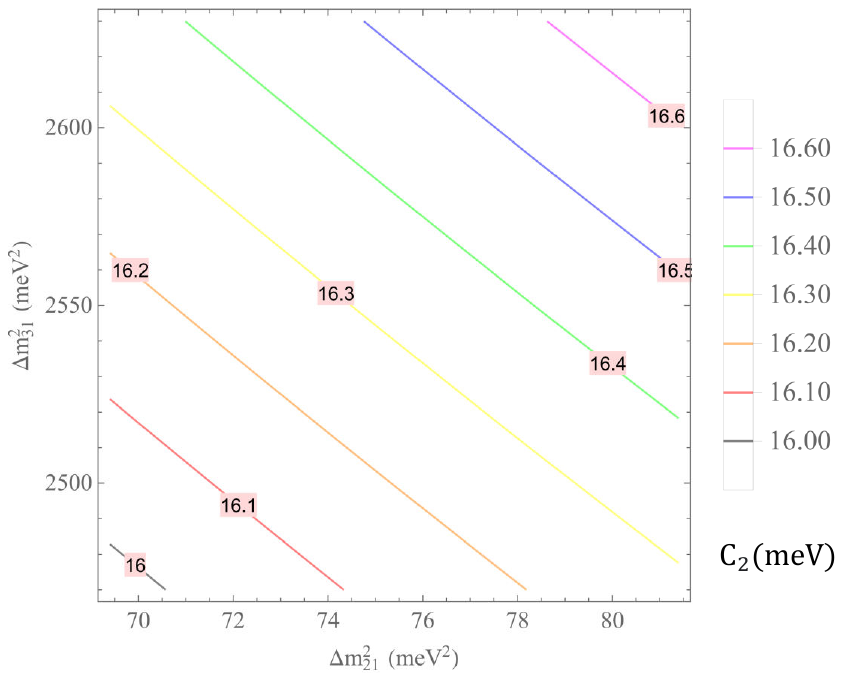}\hspace{-4.15 cm}
\includegraphics[width=0.75\textwidth]{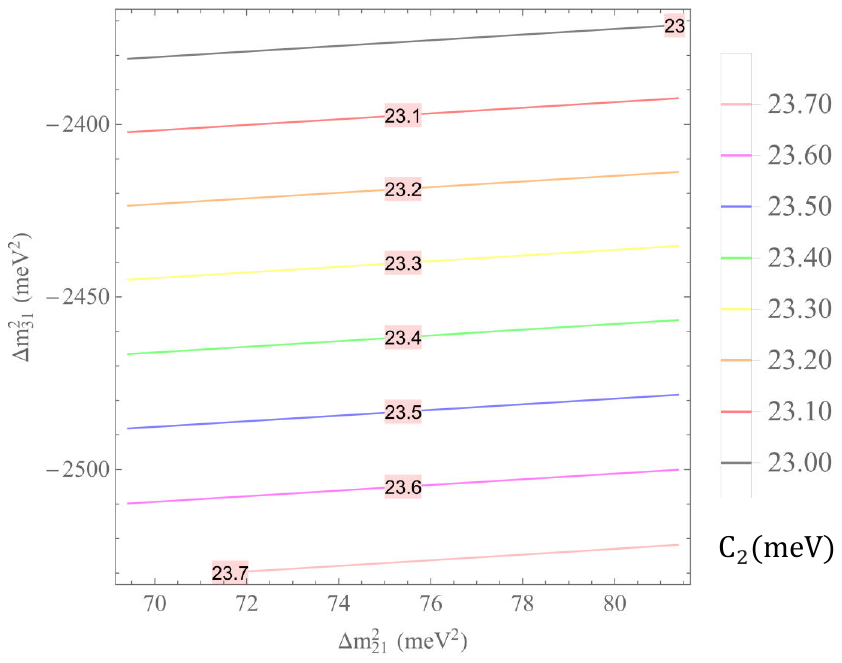}\hspace{-6.25 cm}
\vspace{-8.75 cm}
\caption{$C_2$\,($\mathrm{meV}$) versus $\Delta m^2_{21}$ and $\Delta m^2_{31}$ with $\Delta m^2_{21}\in (69.4, 81.4)\, \mathrm{meV}^2$ and $\Delta m^2_{31}\in (2.47, 2.63) 10^3\, \mathrm{meV}^2$ for NH (left panel) and $\Delta m^2_{31}\in (-2.53, -2.37) 10^3\, \mathrm{meV}^2$ for IH (right panel).}
\label{c2f}
\end{figure}
\vspace{-0.5 cm}
\end{center}
\begin{center}
\begin{figure}[h]
\vspace{-0.5 cm}
\hspace{-5.65 cm}
\includegraphics[width=0.745\textwidth]{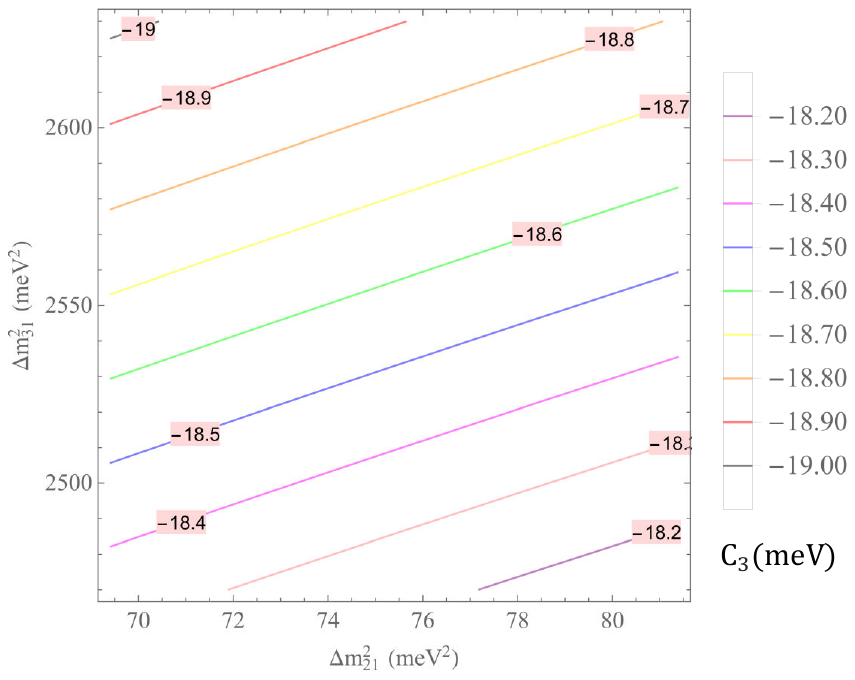}\hspace{-4.1 cm}
\includegraphics[width=0.745\textwidth]{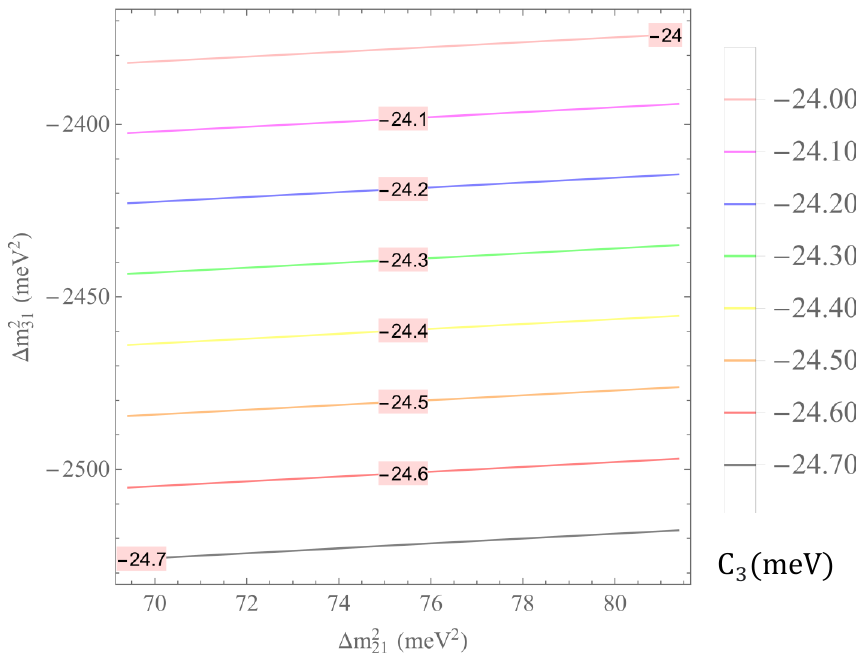}\hspace{-6.25 cm}
\vspace{-8.75 cm}
\caption{$C_3$\,($\mathrm{meV}$) versus $\Delta m^2_{21}$ and $\Delta m^2_{31}$ with $\Delta m^2_{21}\in (69.4, 81.4)\, \mathrm{meV}^2$ and $\Delta m^2_{31}\in (2.47, 2.63) 10^3\, \mathrm{meV}^2$ for NH (left panel) and $\Delta m^2_{31}\in (-2.53, -2.37) 10^3\, \mathrm{meV}^2$ for IH (right panel).}
\label{c3f}
\end{figure}
\vspace{-0.5 cm}
\end{center}
\begin{center}
\begin{figure}[h]
\vspace{-0.75 cm}
\hspace{-5.65 cm}
\includegraphics[width=0.75\textwidth]{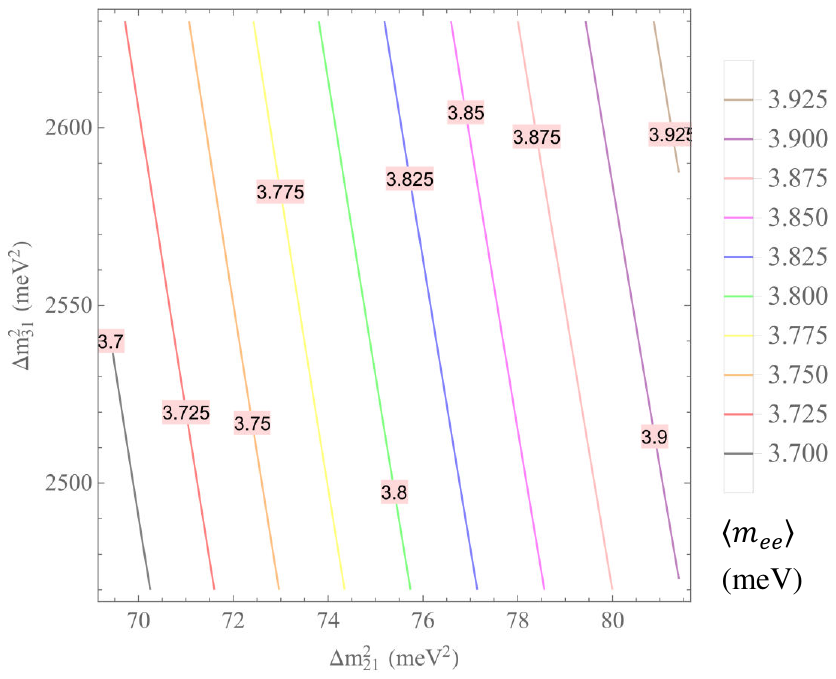}\hspace{-4.25 cm}
\includegraphics[width=0.75\textwidth]{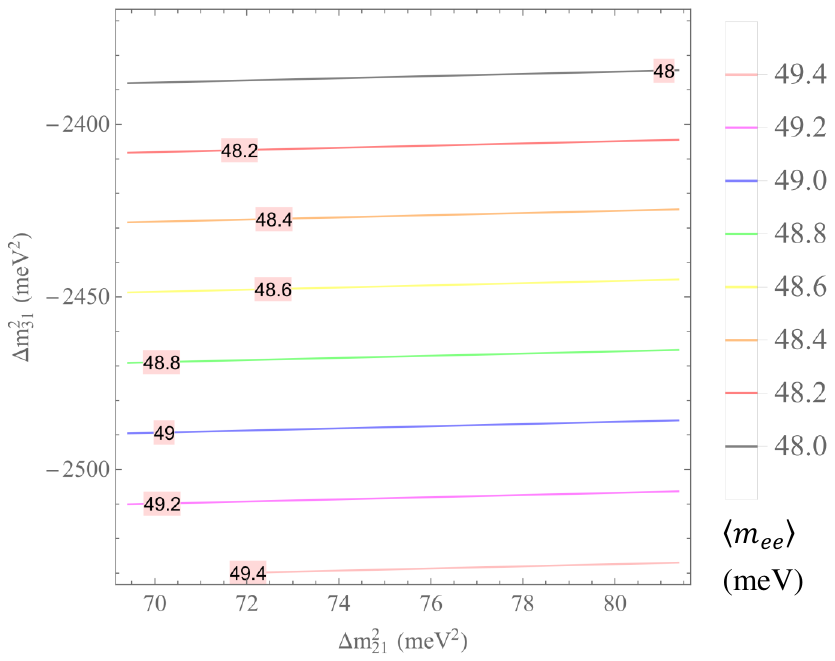}\hspace{-6.25 cm}
\vspace{-8.75 cm}
\caption{$\langle m_{ee}\rangle$\,($\mathrm{meV}$) versus $\Delta m^2_{21}$ and $\Delta m^2_{31}$ with $\Delta m^2_{21}\in (69.4, 81.4)\, \mathrm{meV}^2$ and $\Delta m^2_{31}\in (2.47, 2.63) 10^3\, \mathrm{meV}^2$ for NH (left panel) and $\Delta m^2_{31}\in (-2.53, -2.37) 10^3\, \mathrm{meV}^2$ for IH (right panel).}
\label{meef}
\end{figure}
\vspace{-0.5 cm}
\end{center}
\begin{center}
\begin{figure}[h]
\vspace{-0.65 cm}
\hspace{-5.65 cm}
\includegraphics[width=0.75\textwidth]{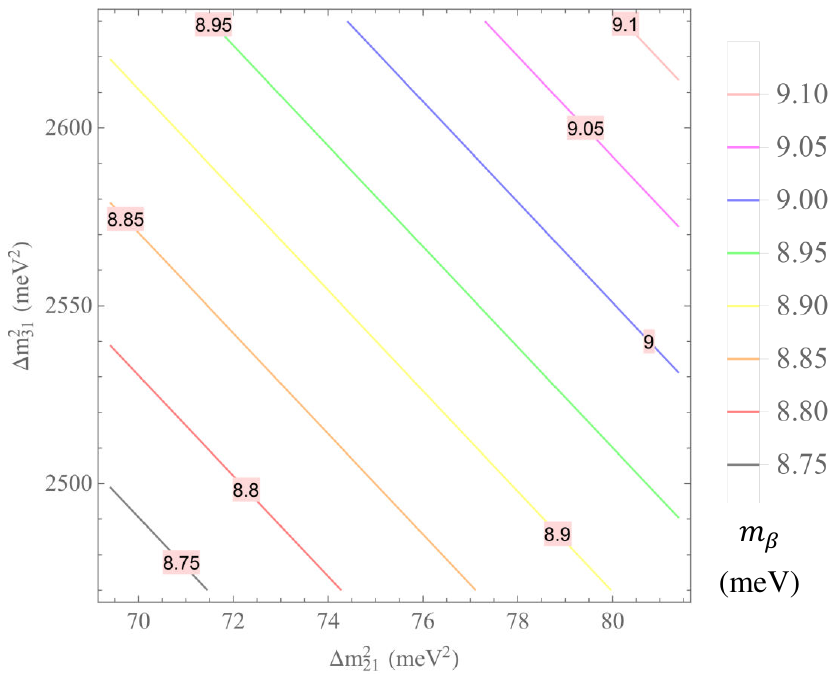}\hspace{-4.25 cm}
\includegraphics[width=0.75\textwidth]{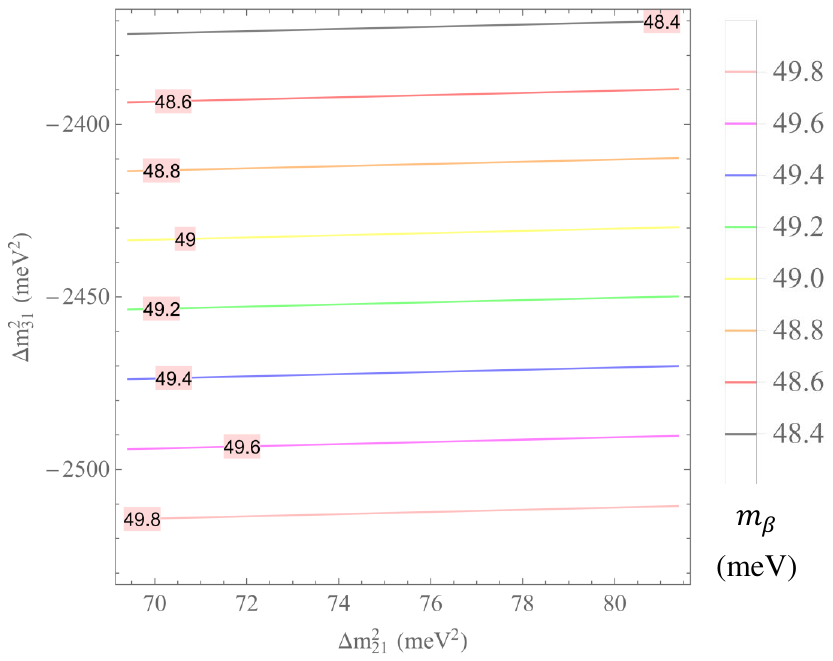}\hspace{-6.25 cm}
\vspace{-8.75 cm}
\caption{$m_{\beta}$\,($\mathrm{meV}$) versus $\Delta m^2_{21}$ and $\Delta m^2_{31}$ with $\Delta m^2_{21}\in (69.4, 81.4)\, \mathrm{meV}^2$ and $\Delta m^2_{31}\in (2.47, 2.63) 10^3\, \mathrm{meV}^2$ for NH (left panel) and $\Delta m^2_{31}\in (-2.53, -2.37) 10^3\, \mathrm{meV}^2$ for IH (right panel).}
\label{mbf}
\end{figure}
\vspace{-0.5 cm}
\end{center}
Figures \ref{af} and \ref{c3f} imply that:
\bea
&&A \in \left\{
\begin{array}{l}
(3.700, 3.925) \, \mathrm{meV}\hspace{0.15cm}\mbox{for  NH},  \\
(48.00, 49.40)\, \mathrm{meV}\hspace{0.15cm}\mbox{for IH},
\end{array}%
\right. \hspace{0.75cm}  B_1\in \left\{
\begin{array}{l}
(3.90, 4.25)\, \mathrm{meV} \hspace{0.5cm}\mbox{for  NH},  \\
(-4.775,-4.60)\, \mathrm{meV} \hspace{0.1 cm}\mbox{for IH},
\end{array}%
\right. \label{c1range}\\
&&B_2\in \left\{
\begin{array}{l}
(-7.15, -6.80)\, \mathrm{meV}\hspace{0.325cm}\mbox{for  NH},  \\
(-5.75, -5.575)\, \mathrm{meV}\hspace{0.1cm}\mbox{for IH},
\end{array}%
\right. \hspace{0.2cm} C_{1}\in \left\{
\begin{array}{l}
(38.40, 39.60)\, \mathrm{meV} \hspace{0.2cm}\mbox{for  NH},  \\
(27.40, 28.20)\, \mathrm{meV} \hspace{0.2cm}\mbox{for IH}.
\end{array}%
\right.\label{sdrange}\\
&&C_2\in \left\{
\begin{array}{l}
(16.00, 16.60)\, \mathrm{meV}\hspace{0.15cm}\mbox{for  NH},  \\
(23.00, 23.70)\, \mathrm{meV}\hspace{0.15cm}\mbox{for IH},
\end{array}%
\right. \hspace{0.65cm} C_{3}\in \left\{
\begin{array}{l}
(-19.00, -18.20)\, \mathrm{meV} \hspace{0.1cm}\mbox{for  NH},  \\
(-24.70, -24.00)\, \mathrm{meV} \hspace{0.1cm}\mbox{for IH}.
\end{array}%
\right.\label{sdrange}
\eea
Figures \ref{meef} and \ref{mbf} show the predictive regions of the effective neutrino-masses:
\bea
&&\langle m_{ee}\rangle \in \left\{
\begin{array}{l}
(3.700, 3.925) \, \mathrm{meV}\hspace{0.15 cm}\mbox{for  NH},  \\
(48.00, 49.40)\, \mathrm{meV}\hspace{0.15 cm}\mbox{for IH},
\end{array}%
\right.  \hs
m_\beta\in \left\{
\begin{array}{l}
(8.75, 9.10)\, \mathrm{meV}\hspace{0.55cm}\mbox{for  NH},  \\
(48.40, 49.80)\, \mathrm{meV}\hspace{0.15cm}\mbox{for IH},
\end{array}%
\right. \label{mbrange}
\eea
which are below the upper limits for $\langle m_{ee}\rangle$ from KamLAND-Zen \cite{KamLAND16} $\langle m_{ee} \rangle <61 \div 165\, \mathrm{meV}$, GERDA \cite{GERDA19} $\langle m_{ee} \rangle < 104\div 228\, \mathrm{meV}$ and CUORE \cite{CUORE20} $\langle m_{ee} \rangle < 75 \div 350 \,\mathrm{meV}$, and the constraints for $m_{\beta}$ with $8.5 \,\mathrm{meV} < m_{\beta} < 1.1\, \mathrm{eV}$ for NH and $48 \, \mathrm{meV} < m_{\beta} < 1.1\, \mathrm{eV}$ for IH \cite{PDG2022}, $m_{\beta} \in (8.90\div 12.60)\, \mathrm{eV}$ \cite{mbet3constraint}, and $m_{\beta} < 0.8\, \mathrm{eV}$ \cite{Aker22n}.
\section{\label{conclusion}Conclusions}
We have constructed a gauge $B-L$ model with $D_4\times Z_4\times Z_2$ symmetry that can explain the quark and lepton mass hierarchies and their mixing patterns with the realistic CP phases via the type-I seesaw mechanism. Six quark mases, three quark mixing angles and CP phase in the quark sector can get the central values 
and Yukawa couplings in the quark sector  are  diluted  a range of three orders of magnitude difference by the perturbation theory at the first order.
For neutrino sector, the smallness of neutrino mass is achieved by the Type-I seesaw mechanism. Both inverted and normal neutrino mass hierarchies are in consistent with the experimental data. The prediction for the sum of neutrino masses is $58. 25 \mathrm{meV} \leq \sum m_\nu \leq 60.25$ meV for normal hierarchy and $98.50 \mathrm{meV}\leq \sum m_\nu \leq 101.00$ meV for inverted hierarchy which are well consistent with all the recent limits. In addition, the Dirac CP phase is predicted to be $288.20\leq \delta (^\circ) \leq 330.00$ within the 3$\sigma$ range of experimental constraint. The effective neutrino masses are predicted to be $3.700\, \mathrm{meV} \leq \langle m_{ee}\rangle \leq 3.925$\, meV, $8.75\, \mathrm{meV} \leq m_{\beta} \leq 9.10\, \mbox{meV}$ for normal hierarchy and $48.00\, \mathrm{meV} \leq \langle m_{ee}\rangle \leq 49.40$ meV and $48.40\, \mathrm{meV} \leq m_{\beta} \leq 49.80\, \mbox{meV}$ for inverted hierarchy which are in consistence with the recent constraints.

\section*{Acknowledgments}
This research is funded by Tay Nguyen University under grant number T2023-45CBT\DH.
\newpage
\appendix
\vspace{-1.25 cm}
\section{Forbidden terms under the model's symmetries}
\begin{table}[h]
\begin{center}
\vspace{-0.75 cm}
\caption{\label{preventedterm} Yukawa terms forbidden by the model's symmetries}
\vspace{0.35 cm}
 \begin{tabular}{|c|c|c|c|} \hline
Yukawa terms & Forbidden by  \\ \hline
$(\overline{\psi}_{\al L} l_{\al R})_{1_{+-}}\widetilde{H}, (\overline{\psi}_{\al L} l_{\al R})_{1_{-+}}\widetilde{H^'};
(\overline{\psi}_{1L} \nu_{\al R})_{2}(H\rho^*)_{2}, (\overline{\psi}_{1L} \nu_{\al R})_{2}(H^'\rho^*)_{2};$&\multirow{7}{1 cm}{\hspace{0.0 cm}$U(1)_Y$}  \\
$(\overline{\psi}_{\al L} \nu_{\al R})_{1_{-+}}(\widetilde{H}\varphi)_{1_{-+}},
(\overline{\psi}_{\al L} \nu_{\al R})_{1_{+-}}(\widetilde{H^'}\varphi)_{1_{+-}};
(Q_{1L}u_{1R})_{1_{++}}(H\phi)_{1_{++}},$&\\
$ (Q_{\al L}u_{\al R})_{1_{++}}(H\phi)_{1_{++}},$
$(Q_{\al L}u_{\al R})_{1_{+-}} H, (Q_{\al L}u_{\al R})_{1_{-+}} H^', (Q_{\al L}u_{\al R})_{1_{--}} (H^'\phi)_{1_{--}};$&\\
$(Q_{1 L}u_{\al R})_{2} (H\rho)_{2}, (Q_{1 L}u_{\al R})_{2} (H^'\rho)_{2},
(Q_{\al L}u_{1 R})_{2} (H\rho^*)_{2}, (Q_{\al L}u_{1 R})_{2} (H^'\rho^*)_{2},$&\\
$(Q_{1L}d_{1R})_{1_{++}}(\widetilde{H}\phi)_{1_{++}}, (Q_{\al L}d_{\al R})_{1_{++}}(\widetilde{H}\phi)_{1_{++}},
(Q_{\al L}d_{\al R})_{1_{+-}} \widetilde{H}, (Q_{\al L} d_{\al R})_{1_{-+}} \widetilde{H^'},$&\\
$  (Q_{\al L}d_{\al R})_{1_{--}} (\widetilde{H^'}\phi)_{1_{--}};
(Q_{1 L}d_{\al R})_{2} (\widetilde{H}\rho)_{2}, (Q_{1 L}d_{\al R})_{2} (\widetilde{H^'}\rho)_{2},$&\\
$(Q_{\al L}d_{1 R})_{2} (\widetilde{H}\rho^*)_{2}, (Q_{\al L}d_{1 R})_{2} (\widetilde{H^'}\rho^*)_{2}$&\\\hline
$(\overline{\nu}^C_{1 R} \nu_{1 R})_{1_{++}} (\phi \chi^*)_{1_{++}},
(\overline{\nu}^C_{1 R} \nu_{1 R})_{1_{++}} (\rho^2)_{1_{++}},
(\overline{\nu}^C_{1 R} \nu_{1 R})_{1_{++}} (\rho^{*2})_{1_{++}}; $&\multirow{3}{1.35 cm}{$U(1)_{B-L}$}  \\
$(\overline{\nu}^C_{\al R} \nu_{\al  R})_{1_{++}} (\phi \chi^*)_{1_{++}},
(\overline{\nu}^C_{\al  R} \nu_{\al  R})_{1_{++}} (\rho^2)_{1_{++}},
(\overline{\nu}^C_{\al  R} \nu_{\al  R})_{1_{++}} (\rho^{*2})_{1_{++}}; $&\\
$(\overline{\nu}^C_{\al  R} \nu_{\al  R})_{1_{+-}} \chi^*; (\overline{\psi}_{1 L} \psi^C_{1 L})_{1_{++}} \widetilde{H^2}, (\overline{\psi}_{1 L} \psi^C_{1 L})_{1_{++}} \widetilde{H^{'2}}.$&\\ \hline
$(\overline{\psi}_{1L} l_{1R})_{1_{++}}H, (\overline{\psi}_{1L} l_{1R})_{1_{++}}H^',
(\overline{\psi}_{1L} l_{1R})_{1_{++}}(H^'\phi)_{1_{--}};
(\overline{\psi}_{1L} \nu_{1R})_{1_{+-}}(H\rho^*)_{2},$&\multirow{6}{1.5 cm}{\hspace{0.5 cm}$D_4$}  \\
$(\overline{\psi}_{1L} \nu_{1R})_{1_{+-}}(H^'\rho^*)_{2};
(\overline{\psi}_{\al L} \nu_{1 R})_{2}(\widetilde{H}\varphi)_{1_{-+}},
(\overline{\psi}_{\al L} \nu_{1 R})_{2}(\widetilde{H^'}\varphi)_{1_{+-}};
(\overline{\nu}^C_{1 R} \nu_{1 R})_{1_{++}} \chi;$&\\
$(\overline{\nu}^C_{1  R} \nu_{\al  R})_{2} \chi, (\overline{\nu}^C_{1  R} \nu_{\al  R})_{2} (\phi\chi)_{1_{++}}; (Q_{1L}u_{1R})_{1_{++}}\widetilde{H}, (Q_{1L}u_{1R})_{1_{++}}\widetilde{H^'}, $&\\
$(Q_{1L}u_{1R})_{1_{++}}(\widetilde{H^'}\phi)_{1_{--}},
(Q_{\al L}u_{\al R})_{1_{++}}\widetilde{H}, (Q_{\al L}u_{\al R})_{1_{++}}\widetilde{H^'}, (Q_{\al L}u_{\al R})_{1_{++}}(\widetilde{H^'}\phi)_{1_{--}}, $& \\
$(Q_{1L}d_{1R})_{1_{++}}H, (Q_{1L}d_{1R})_{1_{++}}H^',(Q_{1L}d_{1R})_{1_{++}}(H^'\phi)_{1_{--}}, $&\\
$(Q_{\al L}d_{\al R})_{1_{++}} H, (Q_{\al L} d_{\al R})_{1_{++}} H^', (Q_{\al L}d_{\al R})_{1_{++}}(H^'\phi)_{1_{--}}$& \\ \hline
$(\overline{\psi}_{1L} \nu_{\al R})_{2}(\widetilde{H}\rho)_{2}, (\overline{\psi}_{1L} \nu_{\al R})_{2}(\widetilde{H^'}\rho)_{2},
(\overline{\psi}_{\al L} \nu_{1 R})_{2}(\widetilde{H}\rho)_{2},
(\overline{\psi}_{\al L} \nu_{1 R})_{2}(\widetilde{H}\rho^*)_{2},$&\multirow{4}{2 cm}{\hspace{0.8 cm}$Z_4$}  \\
$(\overline{\psi}_{\al L} \nu_{1 R})_{2}(\widetilde{H^'}\rho)_{2},
(\overline{\psi}_{\al L} \nu_{1 R})_{2}(\widetilde{H^'}\rho^*)_{2},
(Q_{1 L}u_{\al R})_{2} (H\rho^*)_{2}, (Q_{1 L}u_{\al R})_{2} (H^'\rho^*)_{2},$&\\
$(Q_{\al L}u_{1 R})_{2} (H\rho)_{2}, (Q_{\al L}u_{1 R})_{2} (H^'\rho)_{2};
(Q_{1 L}d_{\al R})_{2} (\widetilde{H}\rho^*)_{2}, (Q_{1 L}d_{\al R})_{2} (\widetilde{H^'}\rho^*)_{2}, $&\\
$(Q_{\al L}d_{1 R})_{2} (\widetilde{H}\rho)_{2}, (Q_{\al L}d_{1 R})_{2} (\widetilde{H^'}\rho)_{2}$&\\
\hline
$(\overline{\psi}_{1L} l_{\al R})_{2}(H\rho)_{2}, (\overline{\psi}_{1L} l_{\al R})_{2}(H^'\rho)_{2},
(\overline{\psi}_{\al L} l_{1 R})_{2}(H\rho^*)_{2}, (\overline{\psi}_{\al L} l_{1 R})_{2}(H^'\rho^*)_{2}$&$Z_2$\\\hline
\end{tabular}
\end{center}
\end{table}
\vspace{-1.0 cm}
\newpage
\section{\label{abcudexpressions} The explicit expressions of $a_{1u,d}, a_{2u,d}, a_{3u,d}, b_{u,d}, c_{1u, d}$, $c_{2u,d}$, $c_{3u}$ and $c_{4u,d}$\\ as functions of quark masses and quark mixing matrix elements}
\vspace{-0.5 cm}
The explicit expressions of $a_{1u,d}, a_{2u,d}, a_{3u,d}, b_{u,d}, c_{1u, d}$, $c_{2u,d}$, $c_{3u}$ and $c_{4u,d}$ are:
\bea
&&a_{1u} = m_u, \hs
a_{2u} = \frac{m_c + m_t}{2}, \hs
a_{3u} =\frac{m_c - m_t}{2},\crn
&&a_{1d} = m_d, \hs
a_{2d} = \frac{m_s+m_b}{2}, \hs
a_{3d} = \frac{m_s-m_b}{2}, \crn
&&c_{1u} = -c_{3u}+\frac{(m_d - m_s) (m_c - m_u) \big(1- \mathrm{V}^{\mathrm{exp}}_{11}\big)}{2 c^*_{1d}},\crn
&&c_{2u}=\frac{(m_{u}-m_{c}) (m_{d}-m_{s})}{c^*_{2d}+c^*_{4d}} \left[\frac{4 b^*_{d} b_{u}}{(m_{b}-m_{s}) (m_{c}-m_{t})}+\frac{c_{4u} (c^*_{2d}+c^*_{4d})}{(m_c-m_u) (m_d-m_s)}+\mathrm{V}^{\mathrm{exp}}_{22}-1\right], \crn
&&c_{3u}=\frac{m_{u}-m_{t}}{2} \left[\frac{(\mathrm{V}^{\mathrm{exp}}_{11}-1) (m_{c}-m_{u}) (m_{d}-m_{s})}{2 c^*_{1d} (m_{t}-m_{u})}+\frac{c^*_{2d}-c^*_{4d}}{m_{b}-m_{d}}+\mathrm{V}^{\mathrm{exp}}_{13}\right], \crn
&&2c_{4u}=\frac{4 b^*_{d} b_{u} (m_{c}-m_{u}) (m_{s}-m_{d})}{(c^*_{2d}+c^*_{4d}) (m_{b}-m_{s}) (m_{c}-m_{t})}-\frac{\mathrm{V}^{\mathrm{exp}}_{33} (m_{b}-m_{d}) (m_{t}-m_{u})}{c^*_{2d}-c^*_{4d}}\crn
&&\hspace{0.85 cm}+\frac{(m_{b}-m_{d}) (m_{t}-m_{u})}{c^*_{2d}-c^*_{4d}}+\frac{\mathrm{V}^{\mathrm{exp}}_{22} (m_{u}-m_{c}) (m_{d}-m_{s})}{c^*_{2d}+c^*_{4d}}+\frac{(m_{c}-m_{u}) (m_{d}-m_{s})}{c^*_{2d}+c^*_{4d}}, \label{abcudexpress}\\
&&b_u=\frac{(m_{b}-m_{s}) (m_{c}-m_{t}) \big\{(c^*_{2d} + c^*_{4d}) \big[2 c^*_{1d} +(m_d - m_s) \mathrm{V}^{\mathrm{exp}}_{21}\big] +(m_d - m_s)^2 \big(1-\mathrm{V}^{\mathrm{exp}}_{22}\big)\big\}}{4 b^*_{d} (m_{d}-m_{s})^2},\hs\crn
&&b^*_{d} = \frac{(m_{b}-m_{d}) (m_{b}-m_{s}) \left\{(1-\mathrm{V}^{\mathrm{exp}}_{11}) (m_{d}-m_{s})^2+2 c^*_{1d} \big[c^*_{2d}+c^*_{4d}+(m_{s}-m_{d}) \mathrm{V}^{\mathrm{exp}}_{12}\big]\right\}}{4 c^*_{1d} (m_{d}-m_{s}) \big[c^*_{4d}-c^*_{2d}+\mathrm{V}^{\mathrm{exp}}_{13} (m_{d}-m_{b})\big]}, \crn
&&c^*_{2d}=\frac{c^*_{4d} \mathrm{V}^{\mathrm{exp}}_{31}+\big(\mathrm{V}^{\mathrm{exp}}_{33}-1\big) (m_{d}-m_{b})}{\mathrm{V}^{\mathrm{exp}}_{31}}, \hs c^*_{4d}=\big\{2 c^*_{1d} (m_{d}-m_{s})\big[m_{b} \mathbf{F}_q+m_{d} \mathbf{G}_q+m_{s} \mathbf{H}_q\big]\crn
&&+(m_{d}-m_{s})^3 \mathbf{T}_q +4 c^{*2}_{1d} (\mathrm{V}^{\mathrm{exp}}_{33}-1) \big[\mathrm{V}^{\mathrm{exp}}_{13} (m_{b}-m_{d})+\mathrm{V}^{\mathrm{exp}}_{12} (m_{d}-m_{s})\big]\big\}/\big\{4 c^*_{1d} \mathrm{V}^{\mathrm{exp}}_{31} \big[2 c^*_{1d} \mathrm{V}^{\mathrm{exp}}_{13}\crn
&&+(m_{d}-m_{s}) (\mathrm{V}^{\mathrm{exp}}_{13} \mathrm{V}^{\mathrm{exp}}_{21}-\mathrm{V}^{\mathrm{exp}}_{23})\big]\big\}, \crn
&&c^*_{1d}=\frac{(m_{s}-m_{d})\sqrt{\mathbf{K}_{1q}}+(m_{s}-m_{d}) \mathbf{P}_{1q}+\mathrm{V}^{\mathrm{exp}}_{13} (m_{d}-m_{s}) \big[(\mathrm{V}^{\mathrm{exp}}_{22}-1) \mathrm{V}^{\mathrm{exp}}_{31}-\mathrm{V}^{\mathrm{exp}}_{21} \mathrm{V}^{\mathrm{exp}}_{32}\big]}{4 \mathrm{V}^{\mathrm{exp}}_{13} \mathrm{V}^{\mathrm{exp}}_{32}-4 \mathrm{V}^{\mathrm{exp}}_{12} \mathrm{V}^{\mathrm{exp}}_{33}}, \nonumber\label{gkudte}
\eea
where
\bea
&&\mathbf{F}_q=(\mathrm{V}^{\mathrm{exp}}_{33}-1) (\mathrm{V}^{\mathrm{exp}}_{13} \mathrm{V}^{\mathrm{exp}}_{21}-\mathrm{V}^{\mathrm{exp}}_{23}), \hs \mathbf{G}_q=\mathrm{V}^{\mathrm{exp}}_{33} \big[\mathrm{V}^{\mathrm{exp}}_{11}+\big(\mathrm{V}^{\mathrm{exp}}_{12}-\mathrm{V}^{\mathrm{exp}}_{13}\big) \mathrm{V}^{\mathrm{exp}}_{21}-\mathrm{V}^{\mathrm{exp}}_{22}+\mathrm{V}^{\mathrm{exp}}_{23}\big]\crn
&&\hspace{0.55 cm}+\mathrm{V}^{\mathrm{exp}}_{22}-\mathrm{V}^{\mathrm{exp}}_{23}-\mathrm{V}^{\mathrm{exp}}_{11}-\mathrm{V}^{\mathrm{exp}}_{12} (\mathrm{V}^{\mathrm{exp}}_{21}+\mathrm{V}^{\mathrm{exp}}_{23} \mathrm{V}^{\mathrm{exp}}_{31})+\mathrm{V}^{\mathrm{exp}}_{13} [\mathrm{V}^{\mathrm{exp}}_{21}+(\mathrm{V}^{\mathrm{exp}}_{22}-1) \mathrm{V}^{\mathrm{exp}}_{31}],\crn
&&\mathbf{H}_q=\big(\mathrm{V}^{\mathrm{exp}}_{11}+\mathrm{V}^{\mathrm{exp}}_{22}\big) (1-\mathrm{V}^{\mathrm{exp}}_{33})+\mathrm{V}^{\mathrm{exp}}_{12} (\mathrm{V}^{\mathrm{exp}}_{21}-\mathrm{V}^{\mathrm{exp}}_{21} \mathrm{V}^{\mathrm{exp}}_{33}+\mathrm{V}^{\mathrm{exp}}_{23} \mathrm{V}^{\mathrm{exp}}_{31})+\mathrm{V}^{\mathrm{exp}}_{13} \mathrm{V}^{\mathrm{exp}}_{31}(1-\mathrm{V}^{\mathrm{exp}}_{22}), \crn
&&\mathbf{T}_q=(1-\mathrm{V}^{\mathrm{exp}}_{11})\big[\mathrm{V}^{\mathrm{exp}}_{21}(1-\mathrm{V}^{\mathrm{exp}}_{33})+\mathrm{V}^{\mathrm{exp}}_{23} \mathrm{V}^{\mathrm{exp}}_{31}\big], \label{FGHTKP}\\
&&\mathbf{K}_{1q}=\big[\big(\mathrm{V}^{\mathrm{exp}}_{11}+\mathrm{V}^{\mathrm{exp}}_{12} \mathrm{V}^{\mathrm{exp}}_{21}\big) \mathrm{V}^{\mathrm{exp}}_{33}-\mathrm{V}^{\mathrm{exp}}_{12} \mathrm{V}^{\mathrm{exp}}_{23} \mathrm{V}^{\mathrm{exp}}_{31}-\mathrm{V}^{\mathrm{exp}}_{13} \mathrm{V}^{\mathrm{exp}}_{21} \mathrm{V}^{\mathrm{exp}}_{32}+\big(\mathrm{V}^{\mathrm{exp}}_{22}-1\big) \mathrm{V}^{\mathrm{exp}}_{13} \mathrm{V}^{\mathrm{exp}}_{31}\crn
&&\hspace{0.70 cm}-\mathrm{V}^{\mathrm{exp}}_{22} \mathrm{V}^{\mathrm{exp}}_{33}+\mathrm{V}^{\mathrm{exp}}_{23} \mathrm{V}^{\mathrm{exp}}_{32}\big]^2+4 \big(\mathrm{V}^{\mathrm{exp}}_{11}-1\big) \big(\mathrm{V}^{\mathrm{exp}}_{13} \mathrm{V}^{\mathrm{exp}}_{32}-\mathrm{V}^{\mathrm{exp}}_{12} \mathrm{V}^{\mathrm{exp}}_{33}\big) \big(\mathrm{V}^{\mathrm{exp}}_{21} \mathrm{V}^{\mathrm{exp}}_{33}-\mathrm{V}^{\mathrm{exp}}_{23} \mathrm{V}^{\mathrm{exp}}_{31}\big), \crn
&&\mathbf{P}_{1q}=(\mathrm{V}^{\mathrm{exp}}_{22}-\mathrm{V}^{\mathrm{exp}}_{11} -\mathrm{V}^{\mathrm{exp}}_{12} \mathrm{V}^{\mathrm{exp}}_{21}) \mathrm{V}^{\mathrm{exp}}_{33}
+(\mathrm{V}^{\mathrm{exp}}_{12} \mathrm{V}^{\mathrm{exp}}_{31}
- \mathrm{V}^{\mathrm{exp}}_{32})\mathrm{V}^{\mathrm{exp}}_{23}.\nonumber
\eea
\newpage
\section{\label{k12n12t12expressions} The explicit expressions of $k_{1,2}, n_{1,2}$ and $t_{1,2}$ as functions of $a_D, b_D, c_D, f_D, g_D, a_R, b_R$ and $c_R$}
The explicit expressions of $k_{1,2}, n_{1,2}$ and $t_{1,2}$ are:
\bea
&&k_1=\frac{c_D - d_D}{a_D + b_D}, \hs k_2=\frac{a_D - b_D}{c_D + d_D}, \\
&&n_1=\Big\{a_D^2(b_D-a_D)(b_R + c_R) - b_D \big[(c_R-b_R)(b_D^2 + 2 c_D d_D) + (c_D^2  +  d_D^2)c_R\big] \crn
&&\hspace{0.5 cm}-a_D \big[b_D^2 (b_R - c_R) + 2 c_D  d_D (b_R + c_R) + (c_D^2 + d_D^2)c_R\big] + (a_D -b_D) \sqrt{\Delta}\Big\}\crn
&&\hspace{0.5 cm}/\Big\{(c_D - d_D) \big\{ b_D^2 (c_R-b_R) + a_D^2 (b_R + c_R) + \big[(c_D + d_D)^2-2 a_D b_D\big]c_R\big\}\Big\}, \\
&&n_2=\Big\{(c_D + d_D) \big\{\big[(a_D+b_D)^2+(c_D-d_D)^2\big] c_R^2 +  (a_D^2-b_D^2) c_R b_R+ 2 (c_D d_D  -a_D b_D) b_R^2\crn
&&\hspace{0.5 cm} - b_R \sqrt{\Delta}\big\}\Big\}/\Big\{(c_D - d_D) \big[b_D^2 b_R (c_R-b_R) + a_D^2 b_R (b_R + c_R) +
   b_R c_R (c_D^2 + d_D^2) - c_R \sqrt{\Delta}\big]\Big\},   \\
   &&t_1=\Big\{a_D^2 (b_D-a_D) (b_R + c_R) +  b_D \big[(b_D^2 + 2 c_D d_D)(b_R - c_R) - (c_D^2+ d_D^2)c_R \big] \crn
&&\hspace{0.5 cm}- a_D \big[b_D^2 (b_R - c_R) + c_D^2 c_R + 2 c_D  d_D (b_R + c_R) + c_R d_D^2\big]
+ (b_D -a_D) \sqrt{\Delta}\Big\}\crn
&&\hspace{0.5 cm}/\Big\{(c_D - d_D) \big[b_D^2 (c_R-b_R) -2 a_D b_D c_R+ a_D^2 (b_R + c_R) +
    c_R (c_D + d_D)^2\big]\Big\} ,\\
   &&t_2=\Big\{(c_D + d_D) \big\{\big[(a_D+ b_D)^2+(c_D-d_D)^2\big]c_R^2 + (a_D^2 -b_D^2) c_R b_R +2 (c_D d_D  - a_D b_D) b_R^2\crn
&&\hspace{0.5 cm}+b_R \sqrt{\Delta}\big]\big\}\Big\}/\Big\{(c_D - d_D) \big[b_D^2 b_R (c_R-b_R) + a_D^2 b_R (b_R + c_R) +
    b_R c_R (c_D^2 + d_D^2) + c_R\sqrt{\Delta}\big]\Big\}, \hspace{0.75 cm}
\eea
where
\bea
&&\Delta=a_D^4 (b_R + c_R)^2 + \big[b_D^2 b_R - (b_D^2 + c_D^2) c_R\big]^2 +
 8 a_D b_D c_D d_D (c_R^2-b_R^2)+
 c_R^2 d_D^4 \crn
&&\hspace{0.5 cm}+
 2 \big[2 b_R^2 c_D^2 - b_D^2 b_R c_R + (b_D ^2- c_D^2) c_R^2\big] d_D^2  + 2 a_D^2 (b_R + c_R) \big[b_D^2 (b_R - c_R) + c_R (c_D^2 + d_D^2)\big]. \hspace{0.5 cm}
\eea

\end{document}